\documentclass{emulateapj}
\usepackage{subfigure}
\shorttitle{SN Ia Hosts and Calibration}

\makeatletter
\newcommand\ionpat[2]{#1$\;${\scshape{#2}}}
\def\sizemlcs17{98.1}
\def\sizesalt2{99.4}
\def\smlcs17partial{94.7}     
\def\ssalt2partial{95.9}
\def\massmlcs17{99.98} 
\def\masssalt2{98.0}
\def\metalmlcs17{91.7}
\def\metalsalt2{57.2}

\def\numbersalt2{77}
\def\numbermlcs17{70}
\def\upperredshift{0.08}
\def\colormlcs17{98.6}
\def\colorsalt2{70}

\def\hubbleconstant{70 km s$^{-1}$ Mpc$^{-1}$}

\def\lowerredshift{0.015}
\def\masslowcosmo{$1+w=0.22^{+0.152}_{-0.108}$}
\def\masshighcosmo{$1+w=-0.03^{+0.217}_{-0.143}$}

\newcommand{\lmneillmassressmarg}{2.1\% (2.0$\sigma$; 40)}

\newcommand{\diffneillmassressmarg}{0.11 (2.1$\sigma$)}
\newcommand{\lmsstrress}{13.8\% (1.1$\sigma$; 26)}

\newcommand{\lmscress}{20.3\% (0.8$\sigma$; 48)}

\newcommand{\lmstwostrresstwo}{2.3\% (2.0$\sigma$; 58)}

\newcommand{\lmstwocresstwo}{13.7\% (1.1$\sigma$; 58)}

\newcommand{\lmmseventeendelresmseventeen}{0.7\% (2.4$\sigma$; 60)}

\newcommand{\lmmseventeenavresmseventeen}{36.4\% (0.4$\sigma$; 60)}

\newcommand{\lmmthirtyoneavresmthirtyone}{0.4\% (2.4$\sigma$; 58)}

\newcommand{\lmlogMhighressmarg}{2.3\% (2.0$\sigma$; 51)}

\newcommand{\difflogMhighressmarg}{0.12 (2.6$\sigma$; 51)}
\newcommand{\lmlogMhighresstwomarg}{0.8\% (2.4$\sigma$; 60)}

\newcommand{\difflogMhighresstwomarg}{0.10 (2.2$\sigma$; 60)}
\newcommand{\lmlogMhighresmseventeenmarg}{1.6\% (2.2$\sigma$; 58)}

\newcommand{\difflogMhighresmseventeenmarg}{0.093 (1.9$\sigma$; 58)}
\newcommand{\lmlogMhighresmthirtyonemarg}{3.2\% (1.8$\sigma$; 56)}

\newcommand{\difflogMhighresmthirtyonemarg}{0.089 (1.6$\sigma$; 56)}
\newcommand{\lmlogMhighresstwo}{0.4\% (3.0$\sigma$; 60)}

\newcommand{\difflogMhighresstwo}{0.11 (2.5$\sigma$; 60)}
\newcommand{\lmlogMhighresmseventeen}{0.2\% (2.8$\sigma$; 58)}

\newcommand{\difflogMhighresmseventeen}{0.11 (2.3$\sigma$; 58)}
\newcommand{\lmlogMhighresmthirtyone}{0.2\% (2.7$\sigma$; 56)}

\newcommand{\difflogMhighresmthirtyone}{0.12 (2.2$\sigma$; 56)}
\newcommand{\lmlogMhighress}{1.1\% (2.3$\sigma$; 51)}

\newcommand{\difflogMhighress}{0.13 (2.8$\sigma$; 51)}
\newcommand{\lmlogMressmarg}{38.4\% (0.3$\sigma$; 53)}

\newcommand{\difflogMressmarg}{0.090 (2.0$\sigma$)}
\newcommand{\lmlogMresstwomarg}{9.8\% (1.2$\sigma$; 62)}

\newcommand{\difflogMresstwomarg}{0.094 (2.1$\sigma$)}
\newcommand{\lmlogMresmseventeenmarg}{5.8\% (1.6$\sigma$; 60)}

\newcommand{\difflogMresmseventeenmarg}{0.083 (1.8$\sigma$)}
\newcommand{\lmlogMresmthirtyonemarg}{11.4\% (1.2$\sigma$; 58)}

\newcommand{\difflogMresmthirtyonemarg}{0.085 (1.5$\sigma$)}
\newcommand{\lmlogMresstwo}{2.3\% (2.0$\sigma$; 62)}

\newcommand{\difflogMresstwo}{0.051 (1.0$\sigma$)}
\newcommand{\lmlogMresmseventeen}{0.4\% (2.4$\sigma$; 60)}

\newcommand{\difflogMresmseventeen}{0.10 (2.2$\sigma$)}
\newcommand{\lmlogMresmthirtyone}{0.7\% (2.4$\sigma$; 58)}

\newcommand{\difflogMresmthirtyone}{0.13 (2.4$\sigma$)}
\newcommand{\lmlogMress}{29.9\% (0.5$\sigma$; 53)}

\newcommand{\difflogMress}{0.11 (2.5$\sigma$)}
\newcommand{\lmneillmassresstwomarg}{0.2\% (2.8$\sigma$; 48)}

\newcommand{\diffneillmassresstwomarg}{0.10 (2.0$\sigma$)}
\newcommand{\lmneillmassresmseventeenmarg}{0.2\% (3.0$\sigma$; 46)}

\newcommand{\diffneillmassresmseventeenmarg}{0.17 (3.2$\sigma$)}
\newcommand{\lmneillmassresmthirtyonemarg}{0.5\% (2.7$\sigma$; 46)}

\newcommand{\diffneillmassresmthirtyonemarg}{0.17 (2.8$\sigma$)}
\newcommand{\lmneillmassresstwo}{0.03\% (3.5$\sigma$; 48)}

\newcommand{\diffneillmassresstwo}{0.11 (2.2$\sigma$)}
\newcommand{\lmneillmassresmseventeen}{0.01\% (3.9$\sigma$; 46)}

\newcommand{\diffneillmassresmseventeen}{0.19 (4.0$\sigma$)}
\newcommand{\lmneillmassresmthirtyone}{0.04\% (3.0$\sigma$; 46)}

\newcommand{\diffneillmassresmthirtyone}{0.21 (3.4$\sigma$)}
\newcommand{\lmneillmassress}{0.7\% (2.5$\sigma$; 40)}

\newcommand{\diffneillmassress}{0.14 (2.7$\sigma$)}
\newcommand{\lmnineradressmarg}{0.6\% (2.4$\sigma$; 53)}

\newcommand{\diffnineradressmarg}{0.082 (1.8$\sigma$; 53)}
\newcommand{\lmnineradresstwomarg}{0.9\% (2.3$\sigma$; 62)}

\newcommand{\diffnineradresstwomarg}{0.10 (2.4$\sigma$; 62)}
\newcommand{\lmnineradresmseventeenmarg}{1.9\% (2.0$\sigma$; 60)}

\newcommand{\diffnineradresmseventeenmarg}{0.089 (1.9$\sigma$; 60)}
\newcommand{\lmnineradresmthirtyonemarg}{2.3\% (1.9$\sigma$; 58)}

\newcommand{\diffnineradresmthirtyonemarg}{0.11 (2.1$\sigma$; 58)}
\newcommand{\lmnineradresstwo}{0.2\% (2.8$\sigma$; 62)}

\newcommand{\diffnineradresstwo}{0.11 (2.6$\sigma$; 62)}
\newcommand{\lmnineradresmseventeen}{0.5\% (2.6$\sigma$; 60)}

\newcommand{\diffnineradresmseventeen}{0.11 (2.3$\sigma$; 60)}
\newcommand{\lmnineradresmthirtyone}{0.4\% (2.7$\sigma$; 58)}

\newcommand{\diffnineradresmthirtyone}{0.17 (3.2$\sigma$; 58)}
\newcommand{\lmnineradress}{0.2\% (2.7$\sigma$; 53)}

\newcommand{\diffnineradress}{0.087 (1.9$\sigma$; 53)}
\newcommand{\lmnineradoutlierressmarg}{9.0\% (1.4$\sigma$; 52)}

\newcommand{\diffnineradoutlierressmarg}{0.083 (1.8$\sigma$)}
\newcommand{\lmnineradoutlierresstwomarg}{5.7\% (1.6$\sigma$; 61)}

\newcommand{\diffnineradoutlierresstwomarg}{0.10 (2.3$\sigma$)}
\newcommand{\lmnineradoutlierresmseventeenmarg}{7.0\% (1.5$\sigma$; 59)}

\newcommand{\diffnineradoutlierresmseventeenmarg}{0.10 (2.2$\sigma$)}
\newcommand{\lmnineradoutlierresmthirtyonemarg}{7.6\% (1.5$\sigma$; 57)}

\newcommand{\diffnineradoutlierresmthirtyonemarg}{0.11 (2.0$\sigma$)}
\newcommand{\lmnineradoutlierresstwo}{2.2\% (2.0$\sigma$; 61)}

\newcommand{\diffnineradoutlierresstwo}{0.11 (2.6$\sigma$)}
\newcommand{\lmnineradoutlierresmseventeen}{1.6\% (2.2$\sigma$; 59)}

\newcommand{\diffnineradoutlierresmseventeen}{0.12 (2.7$\sigma$)}
\newcommand{\lmnineradoutlierresmthirtyone}{1.4\% (2.2$\sigma$; 57)}

\newcommand{\diffnineradoutlierresmthirtyone}{0.17 (3.1$\sigma$)}
\newcommand{\lmnineradoutlierress}{8.0\% (1.4$\sigma$; 52)}

\newcommand{\diffnineradoutlierress}{0.087 (1.9$\sigma$)}

\newcommand{\stdbeforenineradressmarg}{0.168}
\newcommand{\stdafternineradressmarg}{0.156}
\newcommand{\stdbeforenineradresmseventeenmarg}{0.193}
\newcommand{\stdafternineradresmseventeenmarg}{0.184}

\newcommand{\stdbeforenineradoutlierresmthirtyonemarg}{0.219}
\newcommand{\stdafternineradoutlierresmthirtyonemarg}{0.212}
\newcommand{\stdbeforelogMressmarg}{0.168}
\newcommand{\stdafterlogMressmarg}{0.168}
\newcommand{\stdbeforelogMresstwomarg}{0.185}
\newcommand{\stdafterlogMresstwomarg}{0.180}
\newcommand{\stdbeforelogMresmseventeenmarg}{0.193}
\newcommand{\stdafterlogMresmseventeenmarg}{0.183}
\newcommand{\stdbeforelogMresmthirtyonemarg}{0.222}
\newcommand{\stdafterlogMresmthirtyonemarg}{0.211}

\newcommand{\stdbeforelogMresstwo}{0.185}
\newcommand{\stdafterlogMresstwo}{0.178}
\newcommand{\stdbeforelogMresmseventeen}{0.193}
\newcommand{\stdafterlogMresmseventeen}{0.182}
\newcommand{\stdbeforelogMresmthirtyone}{0.222}
\newcommand{\stdafterlogMresmthirtyone}{0.208}
\newcommand{\stdbeforelogMress}{0.168}
\newcommand{\stdafterlogMress}{0.167}
\newcommand{\lmneillmasshighressmarg}{2.0\% (2.0$\sigma$; 40)}

\newcommand{\stdbeforeneillmasshighressmarg}{0.174}
\newcommand{\stdafterneillmasshighressmarg}{0.158}
\newcommand{\diffneillmasshighressmarg}{0.11 (2.1$\sigma$; 40)}
\newcommand{\lmneillmasshighresstwomarg}{1.4\% (2.1$\sigma$; 47)}

\newcommand{\stdbeforeneillmasshighresstwomarg}{0.163}
\newcommand{\stdafterneillmasshighresstwomarg}{0.150}
\newcommand{\diffneillmasshighresstwomarg}{0.083 (1.6$\sigma$; 47)}
\newcommand{\lmneillmasshighresmseventeenmarg}{3.0\% (1.9$\sigma$; 45)}

\newcommand{\stdbeforeneillmasshighresmseventeenmarg}{0.159}
\newcommand{\stdafterneillmasshighresmseventeenmarg}{0.142}
\newcommand{\diffneillmasshighresmseventeenmarg}{0.13 (2.5$\sigma$; 45)}
\newcommand{\lmneillmasshighresmthirtyonemarg}{3.9\% (1.7$\sigma$; 45)}

\newcommand{\stdbeforeneillmasshighresmthirtyonemarg}{0.192}
\newcommand{\stdafterneillmasshighresmthirtyonemarg}{0.180}
\newcommand{\diffneillmasshighresmthirtyonemarg}{0.13 (2.1$\sigma$; 45)}
\newcommand{\lmneillmasshighresstwo}{0.7\% (2.6$\sigma$; 47)}

\newcommand{\stdbeforeneillmasshighresstwo}{0.163}
\newcommand{\stdafterneillmasshighresstwo}{0.150}
\newcommand{\diffneillmasshighresstwo}{0.094 (1.8$\sigma$; 47)}
\newcommand{\lmneillmasshighresmseventeen}{0.4\% (2.8$\sigma$; 45)}

\newcommand{\stdbeforeneillmasshighresmseventeen}{0.159}
\newcommand{\stdafterneillmasshighresmseventeen}{0.143}
\newcommand{\diffneillmasshighresmseventeen}{0.17 (3.2$\sigma$; 45)}
\newcommand{\lmneillmasshighresmthirtyone}{0.7\% (2.4$\sigma$; 45)}

\newcommand{\stdbeforeneillmasshighresmthirtyone}{0.192}
\newcommand{\stdafterneillmasshighresmthirtyone}{0.180}
\newcommand{\diffneillmasshighresmthirtyone}{0.19 (3.0$\sigma$; 45)}
\newcommand{\lmneillmasshighress}{0.3\% (2.7$\sigma$; 40)}

\newcommand{\stdbeforeneillmasshighress}{0.174}
\newcommand{\stdafterneillmasshighress}{0.159}
\newcommand{\diffneillmasshighress}{0.14 (2.7$\sigma$; 40)}
\newcommand{\stdbeforeneillmassressmarg}{0.174}
\newcommand{\stdafterneillmassressmarg}{0.158}
\newcommand{\stdbeforeneillmassresstwomarg}{0.184}
\newcommand{\stdafterneillmassresstwomarg}{0.158}
\newcommand{\stdbeforeneillmassresmseventeenmarg}{0.181}
\newcommand{\stdafterneillmassresmseventeenmarg}{0.151}
\newcommand{\stdbeforeneillmassresmthirtyonemarg}{0.213}
\newcommand{\stdafterneillmassresmthirtyonemarg}{0.188}
\newcommand{\stdbeforeneillmassresstwo}{0.184}
\newcommand{\stdafterneillmassresstwo}{0.158}
\newcommand{\stdbeforeneillmassresmseventeen}{0.181}
\newcommand{\stdafterneillmassresmseventeen}{0.151}
\newcommand{\stdbeforeneillmassresmthirtyone}{0.213}
\newcommand{\stdafterneillmassresmthirtyone}{0.187}
\newcommand{\stdbeforeneillmassress}{0.174}
\newcommand{\stdafterneillmassress}{0.158}
\newcommand{\stdbeforenineradresstwomarg}{0.185}
\newcommand{\stdafternineradresstwomarg}{0.175}
\newcommand{\stdbeforenineradresmthirtyonemarg}{0.222}
\newcommand{\stdafternineradresmthirtyonemarg}{0.210}
\newcommand{\stdbeforenineradresstwo}{0.185}
\newcommand{\stdafternineradresstwo}{0.175}
\newcommand{\stdbeforenineradresmseventeen}{0.193}
\newcommand{\stdafternineradresmseventeen}{0.184}
\newcommand{\stdbeforenineradresmthirtyone}{0.222}
\newcommand{\stdafternineradresmthirtyone}{0.210}
\newcommand{\stdbeforenineradress}{0.168}
\newcommand{\stdafternineradress}{0.156}
\newcommand{\stdbeforenineradoutlierressmarg}{0.161}
\newcommand{\stdafternineradoutlierressmarg}{0.157}
\newcommand{\stdbeforenineradoutlierresstwomarg}{0.182}
\newcommand{\stdafternineradoutlierresstwomarg}{0.177}
\newcommand{\stdbeforenineradoutlierresmseventeenmarg}{0.191}
\newcommand{\stdafternineradoutlierresmseventeenmarg}{0.185}
\newcommand{\stdbeforenineradoutlierresstwo}{0.182}
\newcommand{\stdafternineradoutlierresstwo}{0.177}
\newcommand{\stdbeforenineradoutlierresmseventeen}{0.191}
\newcommand{\stdafternineradoutlierresmseventeen}{0.185}
\newcommand{\stdbeforenineradoutlierresmthirtyone}{0.219}
\newcommand{\stdafternineradoutlierresmthirtyone}{0.212}
\newcommand{\stdbeforenineradoutlierress}{0.161}
\newcommand{\stdafternineradoutlierress}{0.157}

\newcommand{\lmmthirtyonedelresmthirtyone}{3.6\% (1.9$\sigma$; 58)}

\newcommand{\flatresstwomargakaike}{0}
\newcommand{\flatresstwomargbic}{0.0}

\newcommand{\flatressmargakaike}{0}
\newcommand{\flatressmargbic}{0.0}

\newcommand{\flatresmseventeenmargakaike}{0}
\newcommand{\flatresmseventeenmargbic}{0.0}

\newcommand{\flatresmthirtyonemargakaike}{0}
\newcommand{\flatresmthirtyonemargbic}{0.0}

\newcommand{\neillmasshighakresstwomargpartbic}{-2.4}
\newcommand{\neillmasshighakresstwomargpartakaike}{-5.6}

\newcommand{\logmhighakresstwomargpartbic}{-4.5}
\newcommand{\logmhighakresstwomargpartakaike}{-7.7}
\newcommand{\nineradakresstwomargpartbic}{-4.2}
\newcommand{\nineradakresstwomargpartakaike}{-7.4}
\newcommand{\neillmasshighakressmargpartbic}{-3.4}
\newcommand{\neillmasshighakressmargpartakaike}{-6.3}

\newcommand{\logmhighakressmargpartbic}{-3.0}
\newcommand{\logmhighakressmargpartakaike}{-5.8}
\newcommand{\nineradakressmargpartbic}{-7.0}
\newcommand{\nineradakressmargpartakaike}{-9.8}
\newcommand{\neillmasshighakresmseventeenmargpartbic}{-5.7}
\newcommand{\neillmasshighakresmseventeenmargpartakaike}{-8.8}

\newcommand{\logmhighakresmseventeenmargpartbic}{-6.8}
\newcommand{\logmhighakresmseventeenmargpartakaike}{-9.9}
\newcommand{\nineradakresmseventeenmargpartbic}{-7.2}
\newcommand{\nineradakresmseventeenmargpartakaike}{-10.3}
\newcommand{\neillmasshighakresmthirtyonemargpartbic}{-0.7}
\newcommand{\neillmasshighakresmthirtyonemargpartakaike}{-3.8}

\newcommand{\logmhighakresmthirtyonemargpartbic}{-2.7}
\newcommand{\logmhighakresmthirtyonemargpartakaike}{-5.8}
\newcommand{\nineradakresmthirtyonemargpartbic}{-3.2}
\newcommand{\nineradakresmthirtyonemargpartakaike}{-6.3}
\newcommand{\sstrakressbic}{-3.2}
\newcommand{\sstrakressakaike}{-4.6}

\newcommand{\stwostrakresstwobic}{-0.3}
\newcommand{\stwostrakresstwoakaike}{-2.0}

\newcommand{\mseventeendelakresmseventeenbic}{-2.1}
\newcommand{\mseventeendelakresmseventeenakaike}{-3.7}

\newcommand{\mthirtyonedelakresmthirtyonebic}{0.6}
\newcommand{\mthirtyonedelakresmthirtyoneakaike}{-1.0}

\shortauthors{Kelly et al.}
\submitted{Accepted by the Astrophysical Journal}
\begin{document}


\title{Hubble residuals of nearby Type Ia Supernovae are correlated with host galaxy masses}

\email{pkelly3@stanford.edu}
\author{Patrick L. Kelly\altaffilmark{1,2}}
\author{Malcolm Hicken\altaffilmark{3}}
\author{David L. Burke\altaffilmark{1,2}}
\author{Kaisey S. Mandel\altaffilmark{3}}
\author{Robert P. Kirshner\altaffilmark{3}}

\altaffiltext{1}{Kavli Institute for Particle Astrophysics and Cosmology, Stanford University, 382 Via Pueblo Mall, Stanford, CA 94305}
\altaffiltext{2}{SLAC National Accelerator Laboratory, 2575 Sand Hill Rd, Menlo Park, CA 94025}
\altaffiltext{3}{Harvard-Smithsonian Center for Astrophysics, 60 Garden St., Cambridge, MA 02138}

\keywords{supernovae: general}

\begin{abstract}


From Sloan Digital Sky Survey \textit{u'}\textit{g'}\textit{r'}\textit{i'}\textit{z'} imaging, 
we estimate the stellar masses of the host galaxies of 70 low redshift SN Ia (\lowerredshift~$< z < $~\upperredshift) from the hosts' absolute luminosities and mass-to-light ratios.
These nearby SN were discovered largely by searches targeting luminous galaxies, and we find that their host galaxies are substantially more massive than the hosts of SN discovered by the
flux-limited Supernova Legacy Survey. 
Testing four separate light curve fitters, we detect $\sim$2.5$\sigma$ correlations of Hubble residuals with both host galaxy size and stellar mass, such that SN Ia occurring in physically larger, more massive hosts are $\sim$10\% brighter after light curve correction.
The Hubble residual is the deviation of the inferred distance modulus to the SN, calculated from its apparent luminosity and light curve properties, away from the expected
value at the SN redshift.
Marginalizing over linear trends in Hubble residuals with light curve parameters 
shows that the correlations cannot be attributed to a light curve-dependent calibration error. 
Combining 180 higher-redshift ESSENCE, SNLS, and HigherZ SN with 30 nearby SN whose host masses are less than $10^{10.8}$ M$_{\sun}$ in a cosmology fit yields \mbox{\masslowcosmo}, while a combination where the 30 nearby SN instead have host masses greater than $10^{10.8}$ M$_{\sun}$ yields \mbox{\masshighcosmo}.  
Progenitor metallicity, stellar population age, and dust extinction correlate with galaxy mass and may be responsible for these systematic effects. 
Host galaxy measurements will yield improved distances to SN Ia.

\end{abstract}

\maketitle

\section{Introduction}

SN Ia are useful probes of the cosmic expansion history (\citealt{ast06}; \citealt{wv07}; \citealt{rie07}; \citealt{hi09b}; \citealt{kes09}). Their extremely bright explosions can be detected to high redshifts and, after calibration by light curve shape, yield luminosity distances over a wide range of 
cosmic epochs spanning the past 8 Gyr. 
While the peak luminosities of SN Ia vary by a factor of $\sim$3, more slowly-evolving SN Ia are intrinsically brighter \citep{ph93} and bluer \citep{ri96}, a pattern used to calibrate SN Ia and measure luminosity distances to a precision of $\sim$10\%. 
Although initial efforts were limited by small detector sizes \citep{norg89}, search programs using newer, large format cameras discovered that distant SN Ia were $\sim$25\% fainter than expected for any matter-only universe, favoring an accelerating expansion (\citealt{re98}; \citealt{pe99}). 
In this paper, we look for correlations between the Hubble residuals of nearby SN Ia and their host galaxy sizes and masses.







Theoretical models of a carbon-oxygen white dwarf with a mass approaching the Chandrasekhar limit show that it can explode to produce a realistic SN Ia spectrum (\citealt{hill00}; \citealt{kas05}; \citealt{kas07}; \citealt{kas09}). A supersonic deflagration wave, possibly preceded by a more slowly moving carbon fusion flame, synthesizes radioactive $^{56}$Ni during the explosion, and subsequent decays to $^{56}$Co and $^{56}$Fe (\citealt{tr67}; \citealt{co69}) then power the SN light curve. 
How the progenitor accumulates a mass close to the Chandrasekhar limit before the explosion is not yet clear, but plausible scenarios include accretion from a binary companion (\citealt{whe73}; \citealt{han04}) or a merger with a second white dwarf (\citealt{ibe84}; \citealt{web84}). 


Galaxies with little or no star formation can host SN Ia, implying that some fraction of SN Ia have old progenitors.
\citealt{oem79} showed that the SN Ia rate in star-forming galaxies is related to the current rate of star formation, pointing to a second, rapid path to SN Ia. 
Subsequent studies have confirmed an increased SN Ia rate per unit mass in galaxies with younger stellar populations (\citealt{van90}; \citealt{ma05}), which has been used to rule out strongly a simple progenitor population consisting of stars that explode after reaching a single age (\citealt{sul06}; \citealt{pri08}). 
Progenitor populations with detonation delay times including two separate components, `prompt' and `tardy' (\citealt{sca05}), or that form a continuum better fit the SN Ia rate. 
Consistent with a relatively young progenitor population in star-forming hosts,
the distribution of SN Ia in the \textit{g'}-band light of spiral galaxies is similar to that of SN II (Kolmogorov-Smirnov p=0.22; \citealt{kel08}).  
\citealt{ras09} find that core collapse SN (including SN Ib/c and SN II) are more closely associated with brighter \textit{g'}-band pixels than SN Ia when the regions outside of the bulge are considered in grand-design spirals (Kolmogorov-Smirnov p=0.004), although whether SN Ib/c should be included in this comparison is not clear given their preference for chemically enriched environments (\citealt{pri08}) often found in inner, higher surface-brightness regions.

Correlations between Hubble residuals and progenitor properties have the potential to introduce bias into SN Ia cosmological measurements,
because SN at high redshift are likely to have younger and possibly less metal rich progenitors than their low redshift counterparts. 
If correlations are identified, host measurements such as size, mass, or metallicity could be combined with light curve parameters in, for example, a hierarchical Bayesian framework (\citealt{man09}) to improve luminosity distance estimates to SN Ia and control systematics in cosmological measurements. 
As of yet, however, there is no consensus on whether any correlations between host galaxy properties
and SN Ia Hubble residuals exist. 

From spectra of 29 E/S0 galaxies, \citealt{gal08} found that host metallicity correlates with MLCS2k2 Hubble residuals at the 98\% confidence level for a one-sided test. For a two-sided test, in which either positive or negative correlations would be interesting, the \citealt{gal08} confidence would instead be $\sim$96\%, a `two-sided' confidence.  Unlike \citealt{gal08}, \citealt{how09} did not find a trend in Hubble residuals of 57 SN from the flux-limited SNLS 1st-year survey using SiFTO to fit light curves. 
These studies did not attempt to separate host dependence from light curve dependent trends in Hubble residuals, important given the correlation between light curve shape and host galaxy type (i.e. \citealt{ham96}).

Here we look for correlations between the Hubble residuals of nearby (\lowerredshift~$<z<$~\upperredshift) SN Ia and host galaxy properties using four separate light curve fitters. 
We simultaneously fit for
trends in Hubble residuals with both light curve and host galaxy properties to isolate a dependence of the
SN Ia width-color-luminosity relation on host environment. 
In Section \ref{lightcurves}, we briefly discuss the light curve fitters used to analyze SN in this paper. Section \ref{progprop} reviews current understanding of the effect of progenitor properties on SN Ia luminosities and Hubble residuals. The data set and methods used to measure host galaxies sizes, masses and colors are described in Section \ref{data}. We 
present and examine the significance of the trends we find in Hubble residuals with light curve parameters, host galaxy size, and host galaxy stellar masses in Section \ref{results}. Finally, in Section \ref{discuss}, we discuss and interpret these trends.
In this paper, $H_{o}$=\hubbleconstant~is used to calculate galaxy sizes and stellar masses.
A concordance cosmology is assumed (see \citealt{hi09b} for parameter values inferred from cosmology fits).

\section{Light Curve Analysis}
\label{lightcurves}

We use the light curve fits and Hubble residuals from the \citealt{hi09b} cosmology analysis. Their low redshift SN sample was constructed from $\sim$70 existing light curves in the literature and $\sim$130 new light curves (the CfA3 sample; \citealt{hi09a}). CfA3 light curve observations were acquired with the 1.2m telescope at the F. L. Whipple Observatory and reduced using a single pipeline. A high percentage of these SN have spectroscopic (\citealt{ma08}; Blondin et al. 2010, in preparation) and near-infrared photometric data (\citealt{wv08}; Friedman et al. 2010, in preparation). 
Fitters applied to the light curves include: SALT \citep{guy05}, SALT2 \citep{guy07}, and MLCS2k2 \citep{jha07} with $R_{V} = 1.7$ (MLCS17) and $R_{V} = 3.1$ (MLCS31). 


\subsection{MLCS17 and MLCS31}

In the MLCS2k2 \citep{jha07} model, the light curve color, the light curve shape, and the SN intrinsic luminosity are a function of one parameter, $\Delta$. SN that have light curves with low $\Delta$ decay more slowly and are intrinsically brighter and bluer.  MLCS2k2 assumes that all SN Ia follow this $\Delta$ parameterization of light curve width, color, and SN luminosity and that any remaining variation arises from extinction by dust, which both diminishes the apparent brightness of a SN and reddens its color according to the relationship $R_{V} = A_{V}/E(B-V)$ where $A_{V}$ is the extinction and $E(B-V)$ is the color excess. $R_{V}$ parameterizes the wavelength dependence of the dust reddening law.  To fit a light curve, spectral SN Ia templates are first transformed to the observer-frame (as opposed to SN rest-frame) through K-corrections using the procedure in \citealt{nug02}. MLCS2k2 then simultaneously fits for $\Delta$ and the extinction $A_{V}$ using a one-sided exponential prior with scale $\tau_{E(B-V)}=0.138$ for low redshift SN. For higher redshift SN, \citealt{hi09b} uses the `glosz' prior, which incorporates a detection efficiency that depends on redshift, for the ESSENCE SN \citep{wv07} and a modified `glosz' for the Higher-z SN \citep{rie05}.

While the mean Milky Way $R_{V}$ value is 3.1, this value may not accurately describe extinction in other galaxies. \citealt{ast06} and \citealt{con07} find that $\beta$, the coefficient between luminosity and SN color, is close to 2, less than the 4.1 expected if $R_{V}=3.1$ and the SN Ia color variation is due entirely to dust. By comparing average colors of SN Ia at a series of epochs against simulated colors inside the MLCS2k2 framework, \citealt{kes09} measured $R_{V}$ from the SDSS-II SN sample, finding that $R_{V}$ = 2.18$\pm$0.14(stat)$\pm0.48$(syst).  Recognizing this uncertainty, \citealt{hi09b} performed fitting using both $R_{V}=3.1$ and $R_{V}=1.7$, the value that minimizes the scatter in the Hubble diagram, labeling these as MLCS31 and MLCS17. The MLCS2k2 model parameters were fit from many high-quality, low redshift SN including a number of 1991bg-like SN, which are a faint class of SN Ia identified by strong \ionpat{Ti}{II} lines. \citealt{hi09b} note that removing the 1991bg-like SN from MLCS2k2 training sets may yield improved calibration of `normal' SN Ia [see Figure 26 of \citealt{hi09a}].

\subsection{SALT and SALT2}

SALT and SALT2 fit for a stretch parameter ($s$ for SALT, $x_{1}$ for SALT2), a SN color, $c = (B - V)_{t=Bmax} +0.057$, and the time of maximum $B$-band light, $t_{0}$. High stretch SN are more intrinsically luminous and have broad, slowly-decaying light curves.    SALT uses a modified Nugent template (\citeyear{nug02}) to fit each SN light curve and derive light curve parameters.  A 
linear function of $s$ and $c$ gives the distance modulus in the SALT model, including the effects 
of intrinsic color variation and dust reddening in the single $c$ term, 
\begin{equation}
\mu_{B} = m_{B}^{max} -M + \alpha(s-1) - \beta c
\end{equation}
where the coefficients $\alpha$, $\beta$, and $M$ are marginalized over in the cosmological fit. Instead of
using the Nugent template, SALT2 builds a light curve model as a function of stretch and color
from a combination of empirical light curves and spectra from low and high redshifts. This approach benefits from observations of high redshift SN that sample the rest frame UV and creates a model representative of SN light curves at low and high redshift. The stretch term is $x_{1}$ and the color $c$,
\begin{equation}
\mu_{B} = m_{B}^{max} -M + \alpha x_{1} - \beta c
\end{equation}
where the linear coefficients $\alpha$, $\beta$, and $M$ are also marginalized over in the cosmological fit. 
\citealt{hi09b} found $\alpha$ = 1.26 and $\beta$ = 2.87 in the case of SALT and $\alpha$ = 0.109 and $\beta$ = 2.67 for SALT2.
SALT and SALT2 do not include
1991-bg like objects in training.

\section{Progenitor Properties and SN Ia Luminosities}
\label{progprop}

\subsection{Trends in Total SN Luminosities}

\citealt{ham96} noticed that SN Ia with fast-declining light curves and low luminosities are more common in early-type than late-type hosts, the first strong indication that progenitor properties influence the explosion energy. 
Relationships have been found between SN Ia luminosity and the star formation rate in a flux-limited survey from host photometry \citep{sul06}, stellar population age from spectroscopy (\citealt{gal05}; \citealt{gal08}), and stellar population age from host photometry \citep{how09}. However, these studies find similarly strong trends in SN Ia luminosity with host metallicity, leaving open the question of whether stellar age or metallicity is more important (\citealt{gal08}; \citealt{how09}). 
Recently, \citealt{hi09b}  found that the brightest SN Ia occur in intermediate spirals, not the galaxies with the youngest stellar populations.

Galaxies have metallicity gradients such that stars farther from the galaxy center are generally less metal-rich, and studies have examined SN Ia properties at different projected radial offsets (\citealt{wan97}; \citealt{iva00}; \citealt{gal05}; \citealt{boi09}). \citealt{wan97} and \citealt{gal05} found that SN Ia at larger galactocentric radii have less luminous explosions. However, it is likely that this trend would disappear if SN Ia hosted by large elliptical galaxies, which account for
many of the SN Ia at large galactocentric radii, were removed. Large ellipticals host less luminous SN Ia at both small and large galactocentric distances.
\citealt{coop09} recently studied the local galaxy density, the number of galaxies per unit volume, near the hosts of SDSS-II SN Ia. They found that, when SN Ia were detected in blue, star-forming galaxies, a disproportionate percentage of the host galaxies were in regions of low galaxy density, possibly suggesting that prompt SN Ia are more luminous or occur more frequently in 
environments with lower gas-phase metallicities. They found no evidence for an analogous effect among SN Ia
with redder, older host galaxies.


\subsection{Trends in Hubble Residuals}

\citealt{ti03} suggested a plausible mechanism by which progenitor metallicity could affect SN luminosity. Although \citealt{how09} found that variation 
in SN Ia luminosities predicted by the theory could only account for 7-10\% of the observed variation, the effect may alter the SN Ia width-color-luminosity relation sufficiently to produce trends in Hubble residuals (\citealt{kas09}). \citealt{ti03} argue that metal increases the nucleon density in the white dwarf before the explosion, and greater nucleon densities favor the synthesis of neutron-rich, iron-peak elements such as $^{58}$Ni and $^{54}$Fe at the expense of $^{56}$Ni during the burning phase. The diminished $^{56}$Ni mass that results produces a less luminous explosion. 

The \citealt{gal05}  and \citealt{gal08} spectroscopic analyses of  
host galaxy properties constrained trends in Hubble residuals with host galaxy metallicity using MLCS2k2 light curve fits.
\citealt{gal05} analyzed the integrated host spectra of 57 SN Ia taken by panning a spectrograph slit across each galaxy. 
For the 28 galaxies in their sample with ongoing star formation, they calculated metallicities from emission-line diagnostics, finding no strong evidence for a trend in Hubble residuals with gas-phase metallicity. 
In a separate study, \citealt{gal08} measured the metallicities and ages of the stellar populations in E/S0 hosts from absorption features, finding a correlation between metallicity and Hubble residuals with a one-sided confidence level of 98\%. 


\citealt{how09} constrained host metallicities less directly by exploiting the correlation between galaxy mass and metallicity (e.g. \citealt{tre04}; \citealt{gallazzi05}). They fit \textit{u}\textit{g}\textit{r}\textit{i}\textit{z} photometry with the PEGASE2 population synthesis code to find host masses for SN from the flux-limited Supernova Legacy Survey (SNLS) and extrapolated metallicities from stellar mass measurements using the \citealt{tre04} relation. 
They find no statistically significant trend in Hubble residuals with host metallicity, but their result may be consistent with \citealt{gal08} 
(the original paper reversed the reported direction of trend in Hubble residuals; J. Gallagher, private communication). 
Unlike \citealt{gal08} who used the MLCS2k2 fitter, \citealt{how09} analyzed light curves with SiFTO, a fitter implemented by \citep{con08}, who obtained relatively similar cosmological best-fit parameters from SiFTO and SALT2 analyses of SNLS data. 

\citealt{nei09} fit  photometry of 168 low redshift SN host galaxies with the PEGASE2 population synthesis models used by \citealt{how09}. For each host galaxy, they compiled GALEX UV and either Sloan Digital Sky Survey (SDSS) \textit{u'g'r'i'z'} or RC3 Johnson $UBV$ magnitudes. In an effort to select SN Ia with low extinction, they identified  host galaxies that were best-fit by a PEGASE2 model with  $E(B-V) < 0.05$. Of the SN hosted by these galaxies, they selected 22 with SiFTO light curve parameters $s < 0.75$ and $c < 0.7$ to produce a sample with light curve parameters similar to those of intermediate redshift SNLS SN. 
Among these 22 SN, they found a 2.1$\sigma$ trend in SiFTO Hubble residuals with host galaxy age. 
However, they found no correlation between the PEGASE2 extinction and SN Ia color, so
the trend is difficult to interpret. When 
all SN with $s > 0.75$ and $c > 0.7$ were fit together, no correlation between Hubble residuals and host age was seen.

Galaxy morphology is another host property that may correlate with Hubble residuals.
After correction for light curve shape and reddening, \citealt{jha07} found that SN Ia in elliptical hosts were 1$\sigma$ brighter after light curve correction with MLCS2k2 distances, while
\citealt{hi09b} similarly found that SN Ia in E/S0 hosts are brighter after light curve correction than those in Scd/Sd/Irr hosts by 2$\sigma$. 



\section{Data}
\label{data}

The SDSS \citep{yo00} DR 7 \citep{ab08} includes 11663 square degrees of \textit{u'}\textit{g'}\textit{r'}\textit{i'}\textit{z'} imaging taken in 53.9 second exposures with the 2.5 meter telescope in Apache Point, New Mexico. The SDSS filters cover the near-infrared detector sensitivity limit to the ultraviolet atmospheric cutoff \citep{fu96}. Each Sloan frame is a 13.5 x 9.9 arcminute field over a 2048 x 1498 array of 0.396" pixels. 


\subsection{Sample Selection}

Table~\ref{tab:selection} describes the selection of the sample from SN whose light curve fits meet the \citealt{hi09b} `minimal cut' 
criteria for at least one light curve fitter.
The `minimal cuts' are
 $t_{1st} \leq +10$ days and
$\chi_{\nu}^{2} \leq 1.6$ for MLCS17,
and  $t_{1st} \leq +10$ days as well as
$\chi_{\nu}^{2} \leq 1.5$ for MLCS31.
For the SALT sample, \citealt{hi09b} applied the cuts used by \citealt{kow08} to 
construct the Union compilation including
$z \geq 0.015$,
$t_{1st}/s \leq +6$ days,
fit convergence, 
a Hubble residual within 3$\sigma$ of best-fit cosmology, and
five or more separate observations.
The SALT2 cut consists of excluding SN with light curves where $\chi_{\nu}^{2} \geq 10$.

From these SN, we select nearby (\lowerredshift~$<z<$~\upperredshift) SN Ia in the Hubble flow whose host galaxy images
were in the SDSS DR7. A further cut eliminates SN for which all SDSS host galaxy images were contaminated by
SN light. In late-time light curves \citep{lair06}, SN Ia decay to fainter than $\sim$-12 Johnson V magnitudes within one year. To avoid contamination from rising or declining SN, we exclude images of SN hosts taken from three months before to one year after SN detection. We finally select SN without a nearby bright star and those with a MLCS17 light curve fit consistent with 
low extinction $A_{V} < 0.5$ $(R_{V}=1.7)$.

For the MLCS17 and MLCS31 subsamples, we further restrict SN to those with $\Delta < 0.7$, because $0.7 < \Delta < 1.2$ SN were found by \citealt{hi09b} to be poorly calibrated. 60 MLCS17, 58 MLCS31, 53 SALT, and 62 SALT2 SN remain after all of the cuts. 
    
\begin{deluxetable}{lcccccc}
\tablecaption{Sample Selection}
\tablecolumns{6}
\tablehead{\colhead{Criterion}&\colhead{Ia}}
\startdata
\citealt{hi09b}&309&\\
\lowerredshift~$<z<$~\upperredshift&170&\\
SDSS DR 7&89&\\
Residual SN Light&79&\\
Bright Stellar Contamination&76&\\
Low Extinction $A_{V} < 0.5$ $(R_{V}=1.7)$&70&
\enddata
\tablecomments{Number of SN Ia remaining after applying each criterion. (1) SN with light curves fit by \citealt{hi09b} passing the `minimal cuts'; (2) hosts with \lowerredshift~$< z <$ \upperredshift; (3) inside SDSS DR 7 coverage;  (4) hosts observed when expected residual SN light is insignificant; (5) no bright stars superimposed on galaxy or very bright stars nearby (removes SN 2003W, SN 2006bw, and SN 2007O); (6) objects with low extinction.}
\label{tab:selection}
\end{deluxetable}




\section{Methods}

\subsection{Photometry}
To determine SN Ia host galaxy colors and stellar masses, we measure and model host \textit{u'}\textit{g'}\textit{r'}\textit{i'}\textit{z'} magnitudes. SDSS photometry of bright galaxies suffers from overestimated local backgrounds and occasional galaxy `shredding'  by the object detection algorithm [see Fig. 9 of \citealt{bl05}], so we perform new photometry measurements on SDSS images. 
To measure the total galaxy magnitude, we construct a mosaic of nearby frames. A mosaic includes the entire extent of the host light distribution so that the host magnitude can be measured by directly summing photons. To construct each mosaic, we resample SDSS frames to a common grid using SWarp\footnote{http://astromatic.iap.fr/} with the LANCZOS3 kernel, rescaling fluxes according to SDSS zeropoints for each frame. Next, we determine background differences between the adjacent chips by measuring their median difference in their region of overlap. Subtracting away the median, we produce a constant background across the mosaic. 

With SExtractor's (\citealt{be96}) dual-detection mode, we measure the photometric aperture MAG\_AUTO on the \textit{g'}-band image and apply this aperture to the \textit{u'}\textit{g'}\textit{r'}\textit{i'}\textit{z'} pixel-matched mosaic images to measure total galaxy magnitudes on the SDSS data. Because a consistent aperture is applied, galaxy light distributions in lower S/N images will not be inappropriately `shredded.' Low values of the contrast parameter DETECT\_MINCONT correspond to  
aggressive object deblending. We use DETECT\_MINCONT=0.005 to extract host galaxies except for the host of  
SN 2001en, SN 2002jy, and SN 2007bz, where we use 0.03, and the edge-on host of SN 2001V where we use 0.5.
MAG\_AUTO uses a \cite{kron80} radius, and our `total' magnitudes sum flux within 2.5 Kron radii.  The background level of each SDSS mosaic is estimated from the median of polygonal regions placed by hand outside the periphery of each galaxy. 
Galactic extinction values are taken from \citealt{sch98}, and we use KCORRECT (v. 4.13) \citep{bl03} to calculate rest-frame magnitudes. 

\begin{figure}[htp!]
\centering
\subfigure[]{\includegraphics[angle=0,width=3.5in]{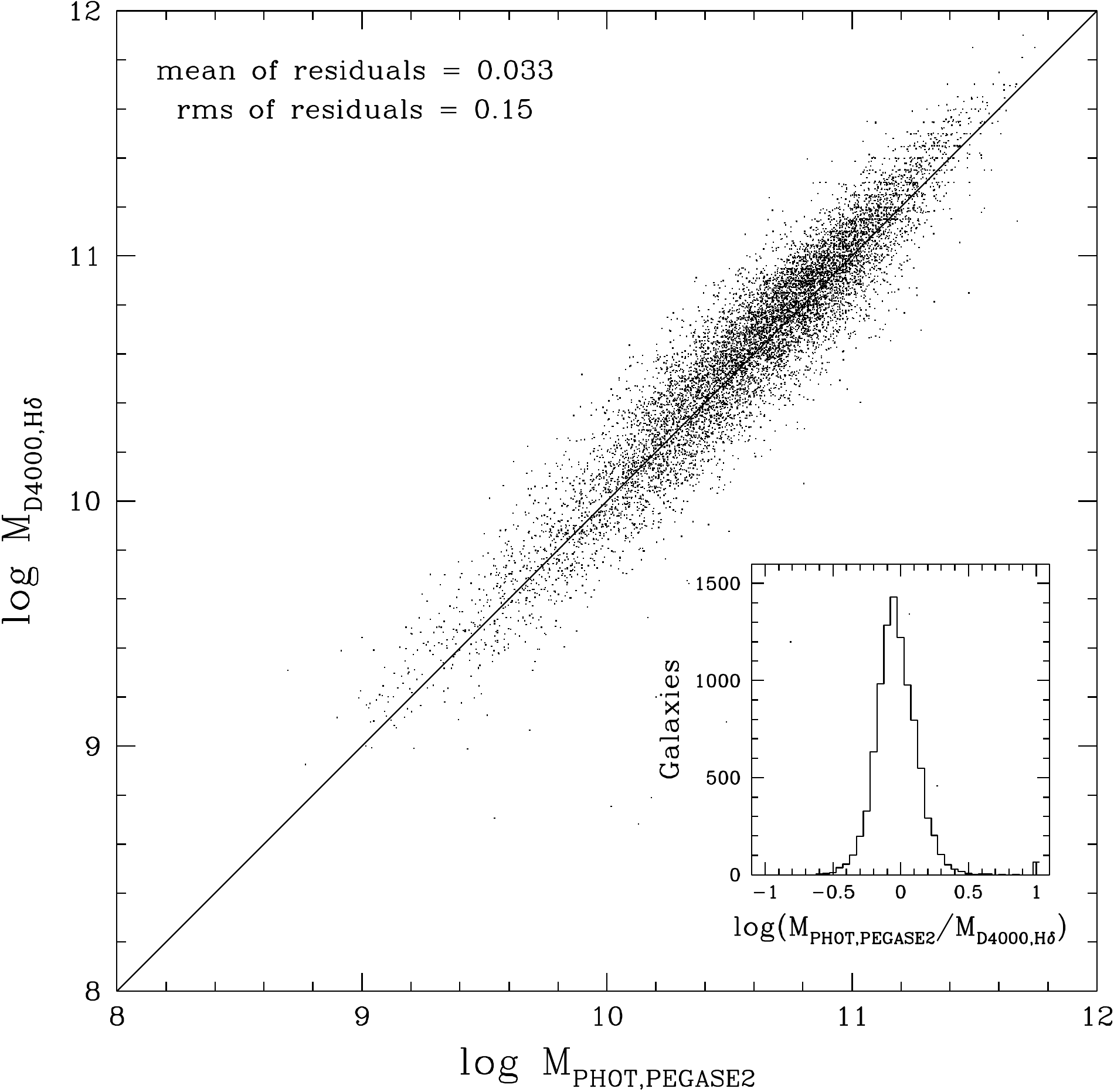}}
\caption{Comparison between our stellar masses from PEGASE2 fits to \textit{u'}\textit{g'}\textit{r'}\textit{i'}\textit{z'} fluxes and \citealt{kau03} stellar masses that use spectral indices stellar Balmer absorption to constrain star formation history. Despite using only \textit{u'}\textit{g'}\textit{r'}\textit{i'}\textit{z'} SDSS photometry (petroMag), we reproduce \citealt{kau03} masses accurately with root-mean-square residuals of 0.15 dex and a mean bias of 0.033 dex. Plotted are 10000 randomly selected $0.05<z<0.20$ galaxies from SDSS DR4 with \citealt{kau03} mass estimates. }
\label{fig:kauffmann}
\end{figure}

\subsection{Galaxy Masses}


\subsubsection{PEGASE2}
\label{pegase}
With the code LePhare (\citealt{arn99}; \citealt{ilb06}), we fit our \textit{u'}\textit{g'}\textit{r'}\textit{i'}\textit{z'} host magnitudes with  spectral energy distributions (SEDs) from PEGASE2 (\citealt{fi97}, \citeyear{fi99}) stellar population synthesis models. 
The SEDs include both nebular and stellar light and track metal enrichment and metal-dependent extinction through 69 time-steps for each of the size models. We use the initial mass function of \citealt{rana92} with a 5\% fraction of
close binaries and a mass range of 0.09 to 120 M$_{\sun}$. Table~\ref{tab:sed} lists the star formation histories used to construct the SEDs. From the marginal probability distribution calculated from the likelihoods of SED fits to galaxy photometry, we find the median stellar mass and uncertainty around the median. 

Near-IR fluxes, which are not included in our fits, have greater sensitivity than optical measurements to the older stellar populations that account for much of galaxies' stellar mass, although the contribution of the asymptotic giant branch remains difficult to model. 
We found, however, that adding 2MASS $JHK$ magnitudes to our fits made only small changes to our stellar mass 
estimates, and we do not include them in our stellar mass fits. 
UV observations may be helpful in the cases of galaxies with both young and old stellar populations to constrain recent star formation history, and
we incorporate into our analysis the available stellar masses from \citealt{nei09} that
were fit to GALEX UV as well as SDSS $\textit{u'}\textit{g'}\textit{r'}\textit{i'}\textit{z'}$ magnitudes.

\citealt{sul06} carefully studied the effects of more complicated star formation histories
as well as uncertain dust extinction on stellar masses estimates calculated using a 
similar set of PEGASE2 SEDs.  
They conclude that the effect of model uncertainties on stellar masses is small
although star formation rates are more strongly affected. 

Table~\ref{tab:all} lists our galaxy measurements. 

\begin{deluxetable}{lccccccc}
\tablecaption{PEGASE2 Star Formation History Scenarios.}
\tablecolumns{4}
\tablehead{\colhead{$\nu$}&\colhead{infall}&\colhead{gal. winds}&\colhead{extinction}&}
\startdata
10&300&300 Myr&spheroidal&\\
2&100&500 Myr&spheroidal&\\
0.66&500&&inclination-averaged&\\
0.4&1000&&inclination-averaged&\\
0.2&1000&&inclination-averaged&\\
0.1&2000&&inclination-averaged&
\enddata
\tablecomments{Summary of PEGASE2 scenarios. SFR=$\nu \times $M$_{gas}$ where $\nu$ has units Gyr$^{-1}$. M$_{gas}$ is the gas density. Infall time scales in Myrs. Spheroidal dust extinction is fitted with a King's profile where the dust density is a power of the stellar density. For inclination-averaged extinction, dust is placed throughout a plane-parallel slab. }
\label{tab:sed}
\end{deluxetable}


\subsubsection{Comparison with Kauffmann Masses}

For 10000 randomly selected SDSS DR4 galaxies, we compare stellar mass estimates calculated from PEGASE2 fits to \textit{u'}\textit{g'}\textit{r'}\textit{i'}\textit{z'} photometry against spectroscopic mass estimates from \citealt{kau03}, which rely on the \citealt{bc03} stellar population synthesis model. 
\citealt{kau03} measure the spectral indices $D_{n}(4000)$, the 4000\AA~break, and  H$\delta_{A}$, stellar Balmer absorption, to constrain star formation history. With a reliable star formation history, internal dust extinction can be estimated by comparing the model prediction to the broadband photometric colors. From star formation histories consistent with the spectral indices and an estimate of internal extinction, \citealt{kau03} find each galaxy's mass-to-light ratio and calculate a mass from the \textit{z'}-band magnitude. 

Figure~\ref{fig:kauffmann} shows a comparison between \citealt{kau03} masses and our mass estimates for 10000 $0.05<z<0.2$ galaxies randomly selected from the SDSS DR4 fit from petroMag catalog magnitudes. Galaxies beyond $z=0.05$ are sufficiently small and faint enough
that the SDSS pipeline photometry is reliable, and the SDSS 3'' fiber samples a significant fraction of the 
galaxy light. The plot shows that the two mass estimates are consistent, with a root-mean-squared dispersion of  0.15 dex and a mean bias of 0.033 dex.

\section{Results}
\label{results}

\subsection{Comparison of Low Redshift SN Host Masses to SNLS 1st-Year Host Masses}

\begin{figure}[htp!]
\centering
\subfigure[]{\includegraphics[angle=0,width=3.5in]{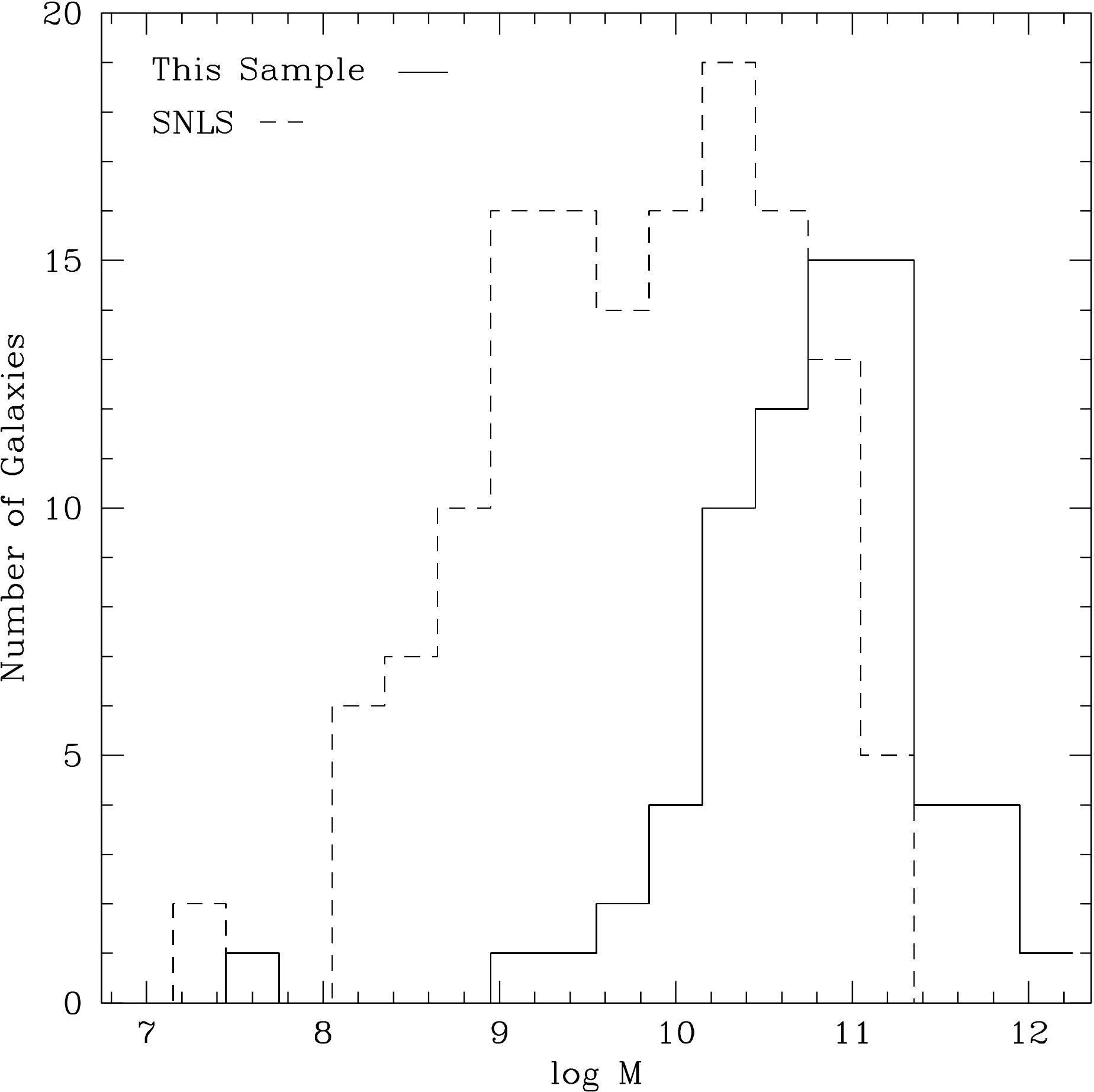}}
\caption{Stellar mass comparison between hosts from this paper's sample and from the SNLS 1st-year survey. Both sets of masses were derived by marginalizing over fits of PEGASE2 stellar population synthesis models to \textit{u}\textit{g}\textit{r}\textit{i}\textit{z} galaxy photometry, although we use a different set of star formation histories that produces better agreement with spectroscopically-determined mass estimates from \citealt{kau03}. These two sets of PEGASE2 templates produce mass estimates with a mean residual offset of 0.009 dex and a root-mean-square spread of 0.12 dex. }
\label{fig:snlsvsus}
\end{figure}

Figure~\ref{fig:snlsvsus} compares host masses of SN in our low redshift sample against those of SN discovered by the Supernova Legacy Survey (SNLS) in its first year as published by \citealt{how09}. The host masses of SNLS SN Ia plotted in Figure~\ref{fig:snlsvsus} were calculated by \citealt{sul06} using PEGASE2 stellar population synthesis models with a somewhat different set of star formation histories than we use to analyze our sample of low-redshift host galaxies. Our choice of star formation histories have a 0.15 dex RMS dispersion in residuals from \citealt{kau03} spectroscopic estimates, while the SNLS masses have a less precise RMS dispersion of 0.17 dex. Nonetheless, SNLS and our mass estimates agree in the mean: after fitting stellar masses to \textit{u'g'r'i'z'} photometry of 10000 randomly selected DR4 galaxies with our set of templates and the SNLS templates, we find a mean offset of 0.009 dex and an RMS of 0.12 dex. Both SNLS and this paper model host galaxy \textit{u}\textit{g}\textit{r}\textit{i}\textit{z} magnitudes, although these correspond to increasingly blue regions of rest-frame galaxy SEDs for more distant SNLS hosts. 

Compared to the SNLS sample, our low redshift host galaxy sample has a much higher percentage of high mass galaxies. This result agrees with 
\citealt{nei09} who fit many of the galaxies in our sample with the PEGASE2 templates used by \citealt{how09}. The high fraction of massive galaxies in our sample is likely largely explained by the fact that many of our \lowerredshift~$<z<$~\upperredshift~SN were discovered by search programs that regularly target luminous, massive local galaxies.  
The Lick Observatory Supernova Search (LOSS), for instance, cycles through a set of host galaxy targets every 3-4 days (\citealt{fili01}).
Of the \numbermlcs17~galaxies in our sample, 17 were discovered by the Lick Observatory Supernova Search (LOSS), 11 were discovered by the Puckett Observatory, and 7 were discovered by the Lick Observatory and Tenegra Observatory Supernova Search (LOTOSS), all of which perform targeted SN searches. 
The SNLS, in contrast, is an untargeted survey which observed four fields during dark and gray time every 3-5 nights for 5-6 lunar phase cycles per year \citep{balland09}. 
Another effect favoring the discovery of SN Ia in massive galaxies at low redshift is the fact that most SN Ia in massive elliptical galaxies are fast-declining. These faint SN are less likely to be discovered at higher redshifts.


Given the correlation between the star formation rate and the SN Ia rate, we expect that the
SN Ia rate will increase at higher redshift where the average star formation rate is higher. Current understanding of star formation histories suggests that 
more massive galaxies form their stars earlier and over a shorter period of time, an effect known as `downsizing' (\citealt{cow96}). 
This suggests that the rate of SN Ia with shorter delay times in more massive galaxies may be elevated at higher redshifts. 
The relatively constant number density of galaxies with masses greater than $10^{10.8}$ M$_{\sun}$ since $z = 0.75$ (i.e. \citealt{poz09})
suggests that an intermediate redshift survey may discover a greater fraction of slow-declining SN Ia in galaxies with masses greater than $10^{10.8}$ M$_{\sun}$ compared to an untargeted low redshift survey.

\subsection{Hubble Residuals and Errors}

In this paper, we define a Hubble residual as \mbox{HR $\equiv$ $\mu_{SN} - \mu_{z}$} so that a positive residual indicates that a SN Ia is fainter after light curve correction than expected for the best-fit cosmology. As calculated by \citealt{hi09b}, the error in HR is the sum in quadrature of errors in $\mu_{SN}$ and $\mu_{z}$.
Error in $\mu_{SN}$ includes uncertainties from the light curve photometry, extinction, the light curve fitter's model coefficients, and an `intrinsic uncertainty' that yields $\chi_{\nu}^{2} \approx 1$ for cosmology fits, while the uncertainty in $\mu_{z}$ comes from peculiar velocity (with a 400 km s$^{-1}$ dispersion). 
Here our sample is a subset of the SN that pass the \citealt{hi09b} `minimal cuts,' and we remove the `intrinsic uncertainty' used by \citealt{hi09b}. 
We add uncertainties that give $\chi_{\nu}^{2} \approx 1$ for the subsamples corresponding to each fitter
(0.14 mag for MLCS17,
0.11 mag for MLCS31,
0.13 mag for SALT,
and 0.13 mag for SALT2). 

We use the 400 km s$^{-1}$ peculiar velocity dispersion adopted by \citealt{hi09b} 
to calculate errors on galaxy properties and the measurement error covariances
between Hubble residuals and galaxy properties. A peculiar velocity towards us makes a SN appear fainter 
than expected for its redshift and also leads to an underestimation of the host galaxy's stellar mass and size.
Conversely, a peculiar velocity away from us makes a SN appear brighter 
than expected for its redshift and also leads to an overestimation of the host galaxy's stellar mass and size, producing measurement error covariances between the SN Hubble residual and host properties.


\begin{figure*}[]
\centering
\subfigure[]{\includegraphics[angle=0,width=3.5in]{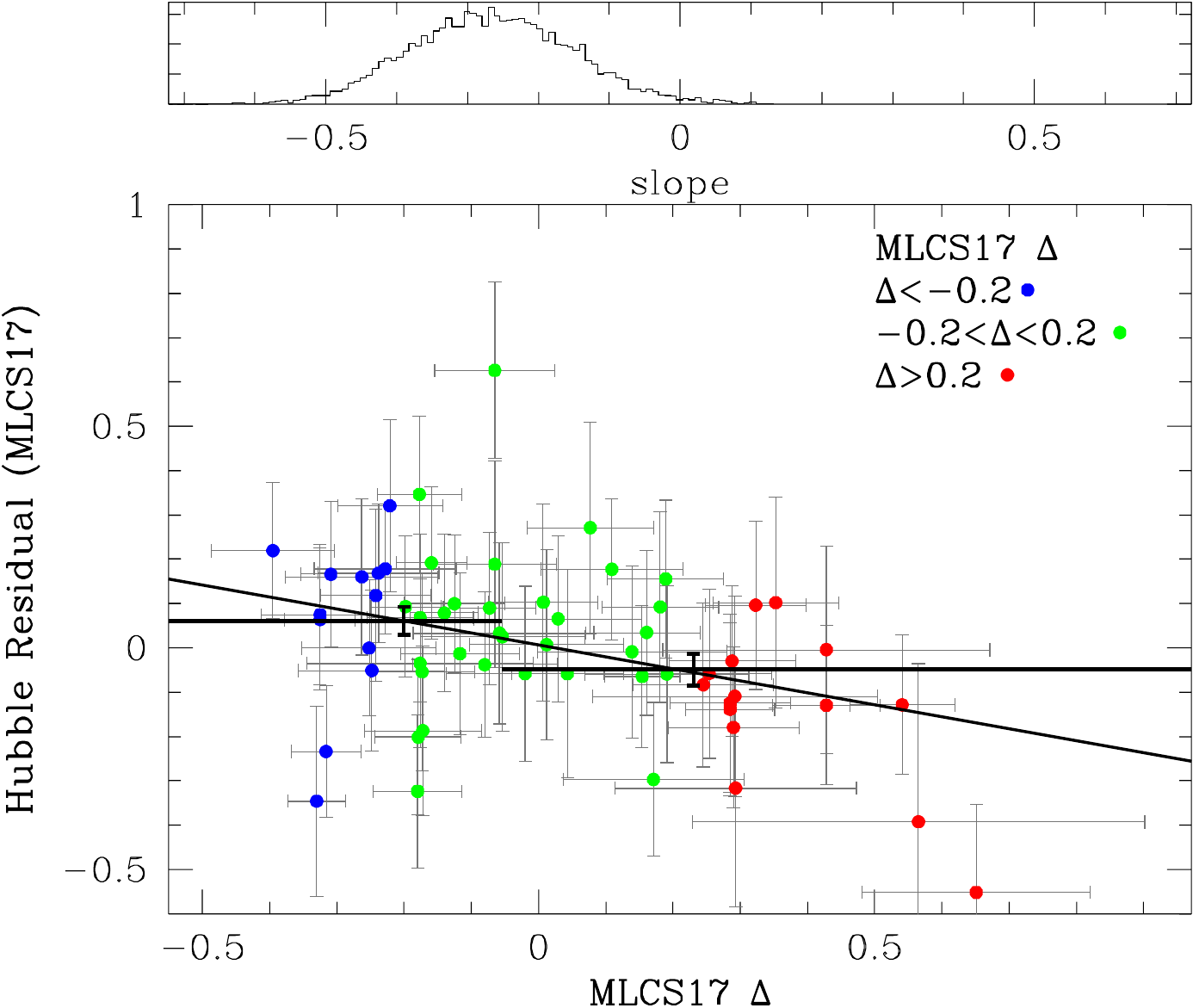}}
\subfigure[]{\includegraphics[angle=0,width=3.5in]{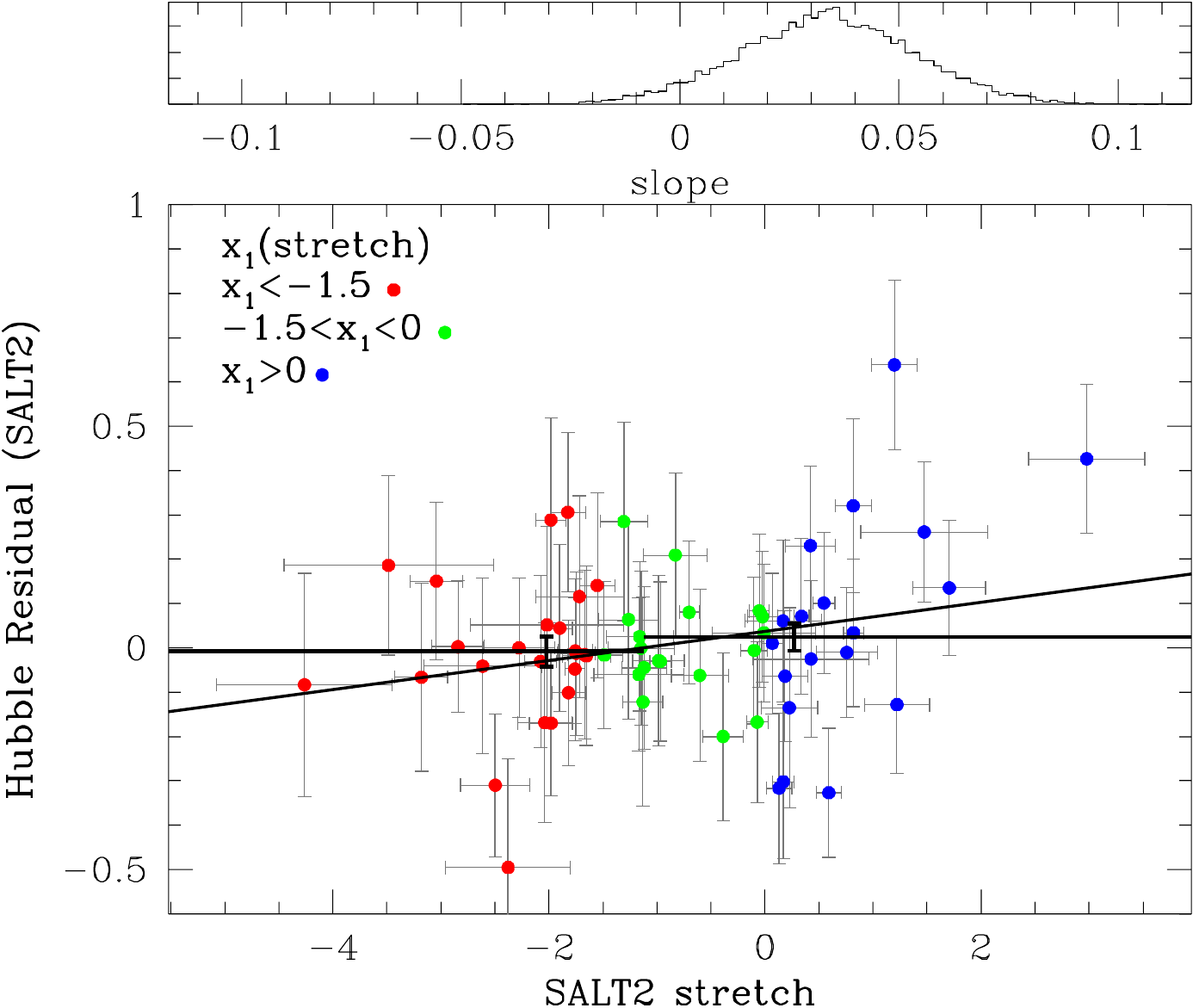}}
\caption{Trends in Hubble residuals with MLCS17 $\Delta$ and SALT2 $x_{1}$. Of our MCMC realizations, \lmmseventeendelresmseventeen~have slopes greater than zero for MLCS17 and  \lmstwostrresstwo~have slopes less than zero for SALT2. 
In parentheses is the significance of a non-zero slope followed by the number of SN included in the fit.
In our analysis of possible trends in Hubble residuals with host properties, we calculate statistics where we marginalize over linear trends with $\Delta$ and stretch. Trends in Hubble residuals with $A_{V}$ or $c$ trend ($\gamma$) for MLCS17 and SALT2 have slopes is consistent with zero. Of our MCMC realizations, \lmmseventeenavresmseventeen~have a slope with $A_{V}$ less than zero for MLCS17 and \lmstwocresstwo~have a slope with $c$ less than zero for SALT2.
}
\label{fig:strtrends}
\end{figure*}

\subsection{Accounting for Calibration Error that Correlates with Light Curve Parameters}


We want to identify trends in Hubble residuals with host galaxy size or mass that cannot be equally well
accounted for by a light curve-dependent calibration error. 
Although host properties and light curve parameters are correlated (i.e. \citealt{ham96}), their correlation is not perfect. The presence of SN Ia with a diversity of light curves in host galaxies within a given size or mass range allows us to disentangle trends in Hubble residuals with light curve parameters from trends with host galaxy properties. 
We fit for linear trends with a host galaxy property (size or mass) and two light curve parameters simultaneously 
so that, for example,
\begin{equation}
\mbox{HR} = \alpha\times(\mbox{host mass}) + \beta\times(\mbox{stretch}) + \gamma\times(\mbox{color}) + \delta
\end{equation}
in the case of the SALT and SALT2 fitters where HR is the Hubble residual.  If the data cannot differentiate between
a Hubble residuals trend with host mass or stretch, for instance, then only the linear combination of $\alpha$ and $\beta$ will be constrained. Consequently, $\alpha$ and $\beta$ can cancel each other out, and they will individually be highly uncertain. If instead a fit finds strong constraints on $\alpha$ and $\beta$, then any trends in host mass and stretch are independent systematic effects in the data. Because host properties and light curve parameters are correlated, we marginalize over uncertainties in the coefficients of trends with light curve parameters when we report uncertainties in the coefficients trends with host size or mass. This marginalization occurs during calculation of posterior probabilities in our Markov Chain Monte Carlo (MCMC) analysis implemented with the LINMIX package \citep{kel07}. Our fits take into account measurement errors as well as measurement error covariance that arises from peculiar velocity.

Here we plot trends in Hubble residuals with light curve parameters to show the reader their size and direction. For brevity, we plot only the trends in MLCS17 Hubble residuals with $\Delta$ and SALT2 residuals with $x_{1}$ in Figure~\ref{fig:strtrends}. Plots of Hubble residuals versus $A_{V}$ for MLCS17 and $c$ for SALT2 are not shown because best-fit trends have slopes consistent with zero. 
The trend in Hubble residuals with MLCS17 $\Delta$, as discussed by \citealt{hi09a}, may reflect the negative quadratic term in $\Delta$ needed to accurately calibrate the quickly decaying, 1991bg-like SN Ia. Removing the 1991bg-like SN from the training set may result in a positive quadratic term which provides a better fit as a function of $\Delta$ for `normal' SN Ia [see Figure 26 of \citealt{hi09a}]. Table~\ref{tab:lightcurve} lists statistics for all four light curve fitters. 

\subsection{Hubble Residuals and Hosts' Physical Sizes}  
\begin{figure*}[]
\centering
\subfigure[]{\includegraphics[angle=0,width=3.5in]{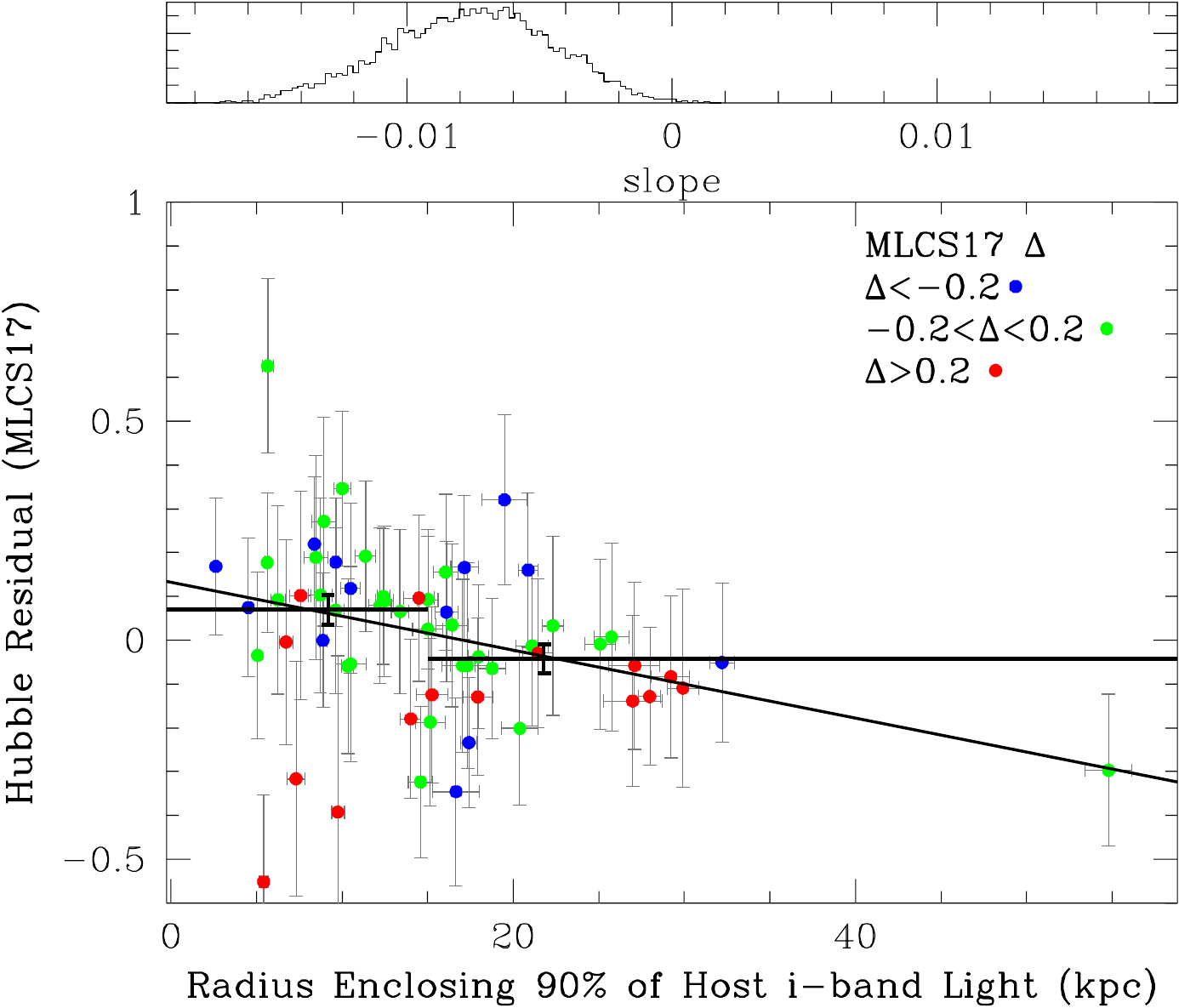}}
\subfigure[]{\includegraphics[angle=0,width=3.5in]{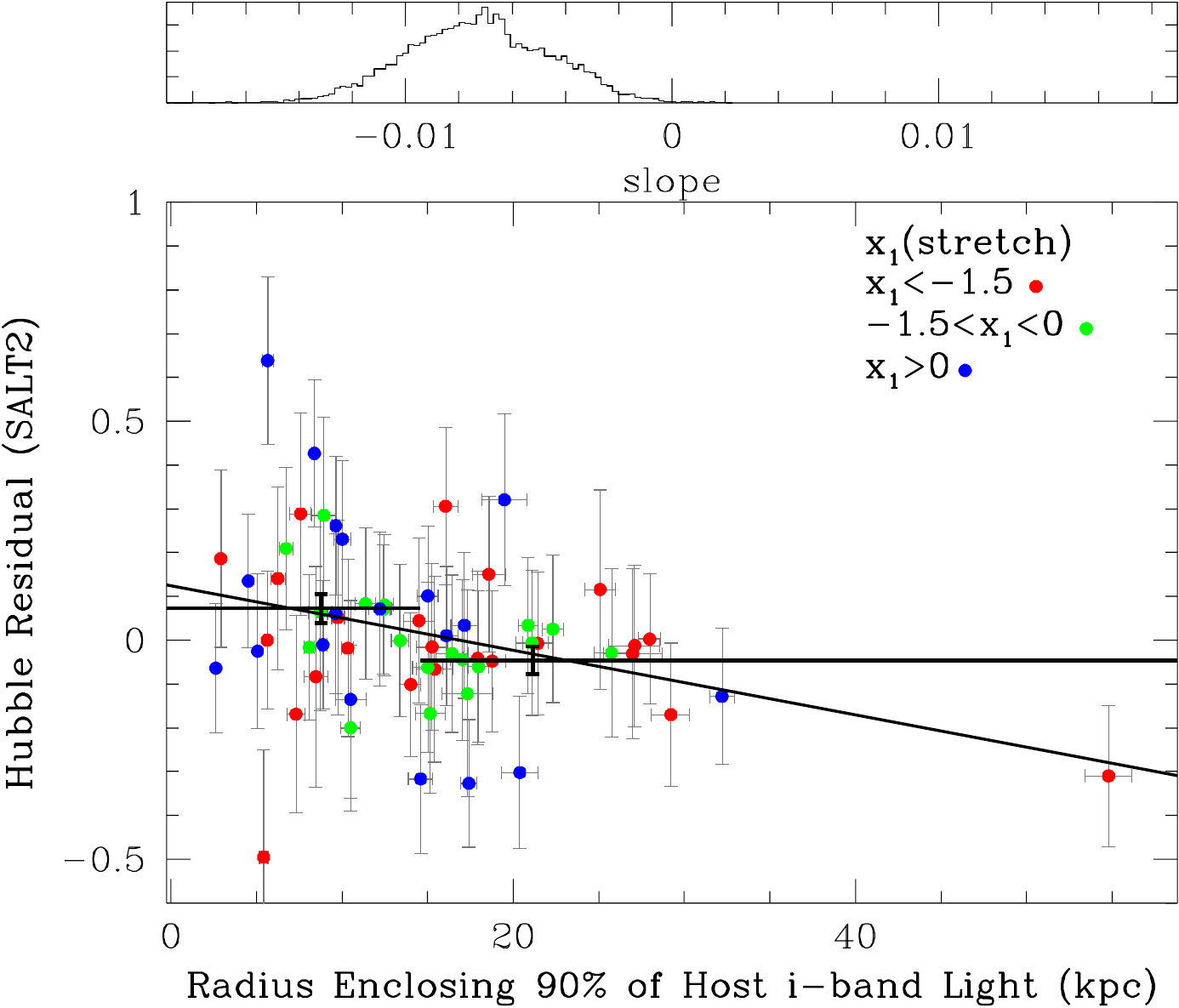}}
\caption{Trends in MLCS17 and SALT2 Hubble residual with host radius enclosing 90\% of \textit{i}-band light. 
SN host radius correlates with Hubble residual for both (a) MLCS17 and (b) SALT2,  with the SN that occur in larger galaxies having significantly more negative Hubble residuals.  The weighted averages of two bins in host size have a magnitude difference of \diffnineradresmseventeen~for MLCS17 and \diffnineradresstwo~for SALT2. When we fit for the trend with host radius ($\alpha$) while holding $\beta$=$\gamma$=0, only \lmnineradresmseventeen~of slopes drawn from an MCMC analysis have greater than zero slope for MLCS17 and \lmnineradresmthirtyone~for SALT2. 
In parentheses is the significance of a non-zero slope followed by the number of SN included in the fit.
Posterior slope distributions are plotted in the upper panels.  The outlying data point is SN 2003ic, hosted by the cD galaxy at the center of the Abell 85 galaxy cluster. With SN 2003ic removed, the bin weighted averages differ in magnitude by \diffnineradoutlierresmseventeen~for MLCS17 and \diffnineradoutlierresstwo~for SALT2 while \lmnineradoutlierresmseventeen~of slopes drawn from an MCMC analysis are greater than zero for MLCS17 and  \lmnineradoutlierresstwo~for SALT2.}
\label{fig:size}
\end{figure*}

We find a trend in Hubble residuals with the radius of the circular aperture enclosing 90\% of the galaxy $\textit{i}$-band light, calculated from SExtractor parameter FLUX\_RADIUS with PHOT\_FLUXFRAC=0.9.  Figure~\ref{fig:size} plots the 90\% host radii against MLCS17 and SALT2 Hubble residuals. The outlying data point with large host radius in the panels is SN 2003ic whose host is the cD galaxy at the center of the galaxy cluster Abell 85. 

First, we fit only for a trend in Hubble residuals with host galaxy size ($\alpha$) while holding $\beta$=$\gamma$=0.  We fit a Gaussian to the posterior probability distribution and use the standard deviation to calculate the $\sigma$ significance of a non-zero slope. 
The significance of a non-zero slope ($\alpha$) ranges  from 2.6-2.9$\sigma$ depending on the light curve fitter. Full statistics are summarized in Tables \ref{tab:1}, \ref{tab:2}, \ref{tab:3}, and \ref{tab:4}.
Separating the SN by host galaxy 90\% radius into two equals-sized bins, the larger-host bin has more negative Hubble residuals according to the bins' weighted averages. The significance of a difference between the bins' weighted averages ranges from 2.3-3.2$\sigma$ among the light curve fitters.

We then simultaneously fit for trends in Hubble residuals with host size ($\alpha$) and light curve parameters ($\beta$,$\gamma$) $\Delta$ and $A_{V}$ for MLCS2k2 fitters and stretch and color for SALT and SALT2. 
Marginalizing over $\beta$ and $\gamma$, the significance of a non-zero slope ($\alpha$) ranges from 1.7-2.4$\sigma$ depending on the light curve fitter (2.0-2.3$\sigma$ excluding MLCS31). The joint posterior probability distributions do not reveal strong degeneracies between the fit coefficients.
Removing the best-fitting trends with light curve parameters ($\beta$,$\gamma$) from the Hubble residuals after the simultaneous fit, a 1.9-2.4$\sigma$ difference between the bin weighted averages persists.  


The cD host galaxy of SN 2003ic is significantly larger than other hosts, and, to test the effect of this data point on our statistics, we repeat the analysis after removing it from the sample. The bin weighted average differences have significances of 1.9-3.1$\sigma$, although a linear fit to the data with a negative slope has only 1.4-2.2$\sigma$ significance (2.0-2.2$\sigma$ without SALT).  

Tables \ref{tab:std} and \ref{tab:stdmarg} show that improvement in the standard deviation of Hubble residuals after removing the host-dependent trend is approximately 5-10\%.

\begin{figure*}
\centering
\subfigure[]{\includegraphics[angle=0,width=3.5in]{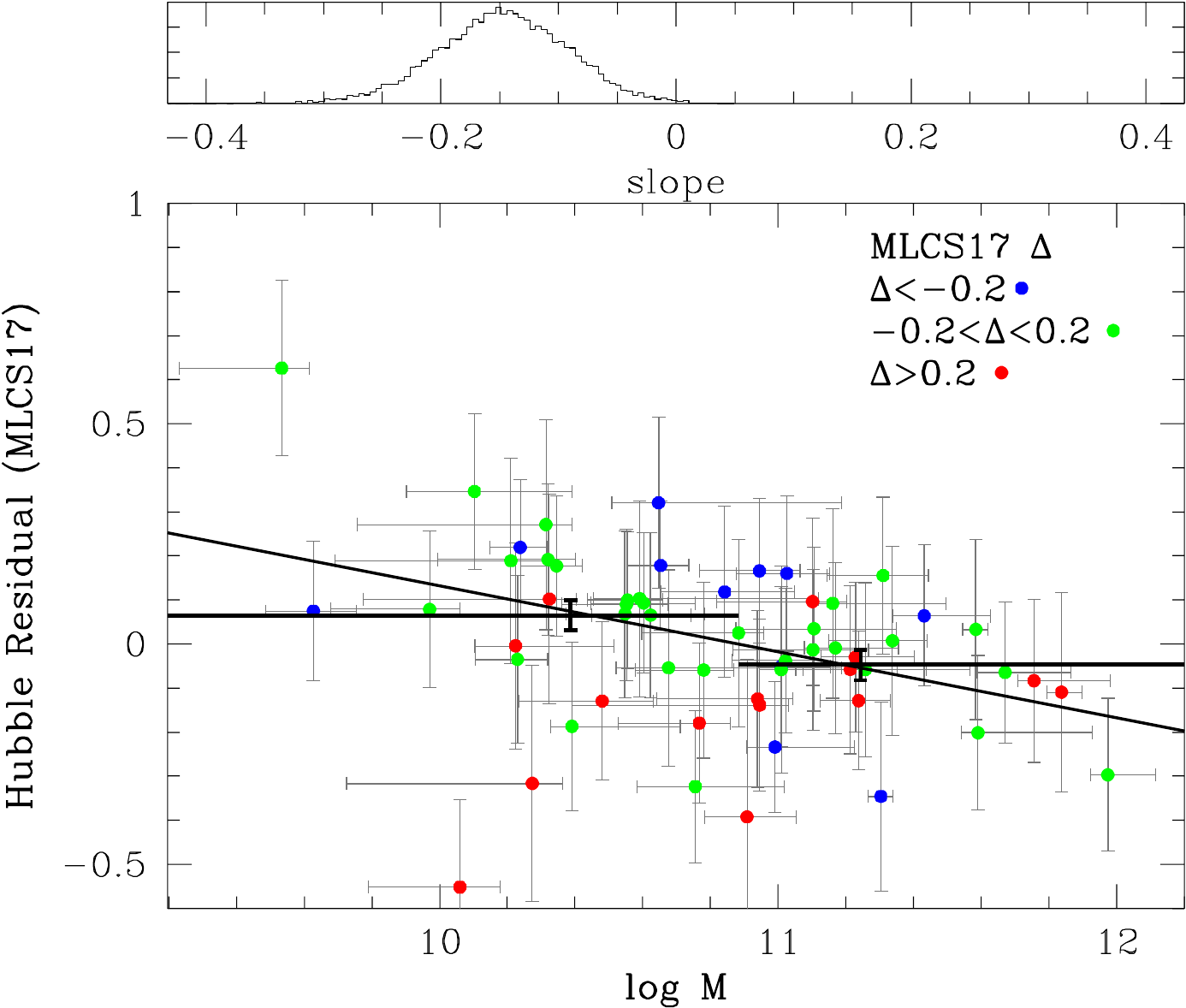}}
\subfigure[]{\includegraphics[angle=0,width=3.5in]{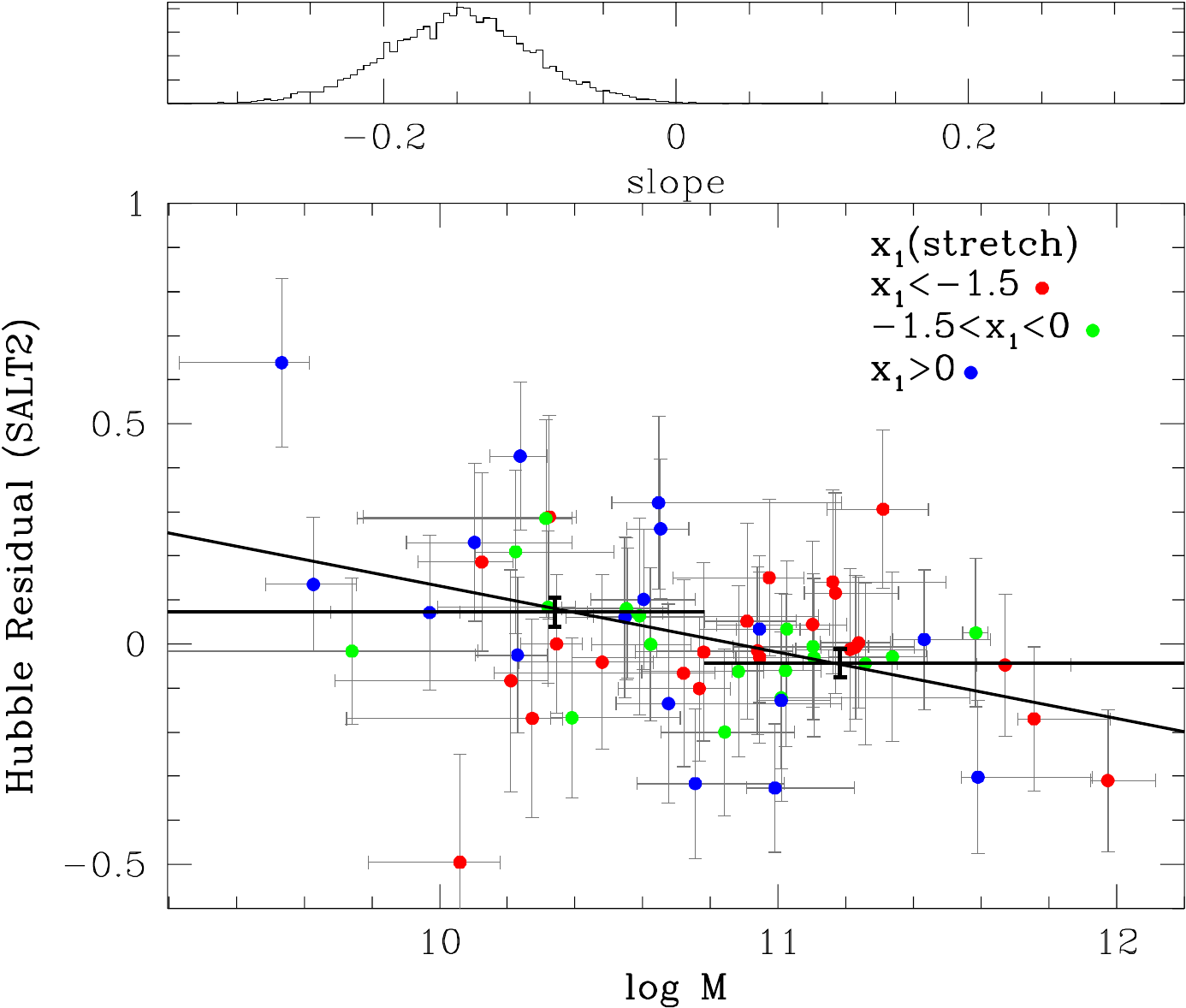}}

\caption{Trends in MLCS17 and SALT2 Hubble residual with host stellar mass (log M$>9.5$). SN with more massive hosts have Hubble residuals that are more negative in the cases of (a) MLCS17 and (b) SALT2, consistent with the trend evident in Figure~\ref{fig:size}. Host stellar masses were measured using fits to \textit{u}\textit{g}\textit{r}\textit{i}\textit{z} photometry with the PEGASE2 stellar population synthesis templates.  The weighted averages of two bins separated by their masses yield a magnitude difference of \difflogMhighresmseventeen~for MLCS17 and \difflogMhighresstwo~for SALT2.  When we fit for the trend with host stellar mass ($\alpha$) while holding $\beta$=$\gamma$=0, only \lmlogMhighresmseventeen~of slopes drawn from a MCMC analysis are greater than zero for MLCS17 and \lmlogMhighresstwo~for SALT2. 
In parentheses is the significance of a non-zero slope followed by the number of SN included in the fit.
The upper panels plot the posterior slope distributions. 
}
\label{fig:mass}
\end{figure*}

\subsection{Hubble Residuals and Galaxy Masses}
Mass is a fundamental property of galaxies which helps determine their star formation histories and chemical enrichment.  Spectroscopic metallicity measurements have shown that stellar mass correlates with galaxy gas-phase \citep{tre04} and stellar \citep{gallazzi05} metal content, reflecting the fact that it is easier for metals to escape from the shallower gravitational potentials of less massive galaxies. As seen in Figure \ref{fig:snlsvsus}, only 2 of our SN host galaxies have masses less than $10^{9.5}$ M$_{\sun}$. 
\citealt{ti03} argue that the effect of metallicity on SN Ia luminosity depends non-linearly on the progenitor's metallicity, which might mean that any effect on Hubble residuals would only become strong for SN Ia in massive galaxies. 
We restrict our sample for our primary analysis to SN with galaxy hosts more massive than $10^{9.5}$ M$_{\sun}$ where we have a sufficient number of SN with massive hosts to constrain any systematic trend.

We plot galaxy stellar mass against Hubble residuals in Figure~\ref{fig:mass} for the MLCS17 and SALT2 light curve fitters, excluding the 2 SN with the least massive hosts. Tables~\ref{tab:1}, \ref{tab:2}, \ref{tab:3}, and \ref{tab:4} summarize all of our statistical tests. 
First, we fit only for a trend in Hubble residuals with host galaxy mass ($\alpha$) while holding $\beta$=$\gamma$=0.
The significance of a non-zero slope ($\alpha$) ranges from 2.3-2.8$\sigma$ depending on the light curve fitter. 
Separating the SN by host galaxy mass into two equal-sized bins, the more massive host bin has more negative Hubble residuals according to the bins' weighted averages. The significance of a difference between the bins' weighted averages ranges from 2.2-2.8$\sigma$ among the light curve fitters.

We then simultaneously fit for trends in Hubble residuals with host mass ($\alpha$) and light curve parameters ($\beta$,$\gamma$) $\Delta$ and $A_{V}$ for MLCS2k2 fitters and stretch and color for SALT and SALT2. 
Marginalizing over $\beta$ and $\gamma$, the significance of a non-zero slope ($\alpha$) ranges from 2.0-2.6$\sigma$ depending on the light curve fitter. 
The joint posterior probability distributions do not reveal strong degeneracies between the fit coefficients.
Removing the best-fitting trends with light curve parameters ($\beta$,$\gamma$) from the Hubble residuals after the simultaneous fit, a 1.6-2.6$\sigma$ difference between the bin weighted averages persists.  

Tables \ref{tab:std} and \ref{tab:stdmarg} show that the reduction in the standard deviation of Hubble residuals after removing the host-dependent trend is approximately
5-10\%.

\subsubsection{\citealt{nei09} Masses}

We now repeat the analyses using stellar mass estimates from \citealt{nei09}, which 
were fit using GALEX UV measurements in addition to SDSS \textit{u'}\textit{g'}\textit{r'}\textit{i'}\textit{z'} 
magnitudes. Because each \citealt{nei09} host galaxy is required to have UV as well as optical measurements, \citealt{nei09} measured masses for only 49 of the 70 SN in our sample with MLCS17 $A_{V} <  0.5$. 
Fitting only for a trend in Hubble residuals with host galaxy mass ($\alpha$) while holding $\beta$=$\gamma$=0,
the significance of a non-zero slope ($\alpha$) ranges from 2.3-2.7$\sigma$ depending on the light curve fitter. 
The significance of a difference between the bins' weighted averages ranges from 1.8-3.2$\sigma$ among the light curve fitters.

Simultaneously fitting for and marginalizing over linear trends ($\beta$,$\gamma$) with light curve parameters,
the significance of a non-zero slope ($\alpha$) is 1.9$\sigma$ for all light curve fitters. 
The joint posterior probability distributions do not reveal strong degeneracies between the fit coefficients.
Removing the best-fitting trends with light curve parameters ($\beta$,$\gamma$) from the Hubble residuals after the simultaneous fit, a 1.6-2.5$\sigma$ (2.1-2.5$\sigma$ without SALT2) difference between the bin weighted averages remains.  

Tables \ref{tab:std} and \ref{tab:stdmarg} show the improvement in the standard deviation of Hubble residuals after removing the host-dependent trend.
This results in improvements of up to ~15\% in standard deviation.

\subsection{Information Criteria}

The Bayesian information criterion (BIC) and the Akaike information criterion (AIC) help to evaluate 
whether an additional model parameter 
can be justified by the improvement it produces in the 
$\chi^{2}$ goodness-of-fit statistic.
BIC derives from an approximation to the Bayes factor, the ratio
of the probabilities of two models given equal prior probability.
For any two prospective models, the model with 
the lower value of BIC is the preferred one. 
With Gaussian errors, the BIC (\citealt{sch78}) is given as,
\begin{equation}
\mbox{BIC} = \chi^{2} + k\ln N
\end{equation}
where $k$ is the number of parameters and $N$ is the number of data points in the fit. 
A comparison can be made between two models by computing
\mbox{$\Delta\mbox{BIC} = \Delta\chi^{2}+\Delta k\ln N$}. 
The AIC \citep{aka74} uses Kullback-Leibler information entropy as a basis to compare different models of a given data set.
The AIC$_{c}$ is a version of the AIC corrected for small data sets \citep{sug78},
\begin{equation}
\mbox{AIC}_{c} = \chi^{2} + 2k + \frac{2k(k + 1)}{N-k-1}
\end{equation}
where $k$ is the number of parameters and $N$ is the number of data points used in fit. 
A decrease of 2 in the information criterion is positive evidence against the model with higher
information criterion, while a decrease of 6 is strong positive evidence against the model with larger value (e.g. \citealt{kas95}, \citealt{muk98}).
To calculate these statistics, we first add an `intrinsic uncertainty' in quadrature to the Hubble 
residual error on each SN so that $\chi^{2}_{\nu}$ = 1 for a fit with no trends. 
We calculate the BIC and AIC$_{c}$ for three separate models:
no trend ($\delta$), a trend with light curve shape parameter ($\beta,\delta$), and
trends with light curve shape parameter and host galaxy property ($\alpha,\beta,\delta$).
Only differences in information criteria are significant, and  Tables \ref{tab:bic} and \ref{tab:aic} 
list the change in the information criteria relative to the 
model with no trend ($\delta$).
An $A_{V}/c$ trend ($\gamma$) is not included because its slope is consistent with zero and its inclusion results in no significant reduction in the fit $\chi^{2}$. 

We are interested in the changes in the BIC and AIC$_{c}$ after adding a host dependent trend ($\alpha$) 
to a model that already includes a trend with light curve shape ($\beta,\delta$). For all light curve fitters, an added trend with host size produces $\Delta\mbox{BIC} < -2$, providing evidence in favor of the additional trend. 
For trends with host stellar mass, $\Delta\mbox{BIC} < -2$ except for SALT. MLCS17 and SALT2 have 
$\Delta\mbox{BIC} < -2$ for \citealt{nei09} stellar masses. 
Only if $\Delta\mbox{BIC} > 2$ would there be evidence against adding a host dependent term, and in no case is $\Delta\mbox{BIC} > 2$.
$\Delta$AIC$_{c}$ is more negative in all cases than $\Delta$BIC 
because BIC has a stronger penalty for each additional parameter. In the cases of all trends and all light curve 
fitters, $\Delta\mbox{AIC}_{c} < -2$ and largely is less than -4, providing moderate to strong positive evidence
for a host dependent term.




\begin{figure*}[htp!]
\centering
\subfigure[]{\includegraphics[angle=0,width=3.2in]{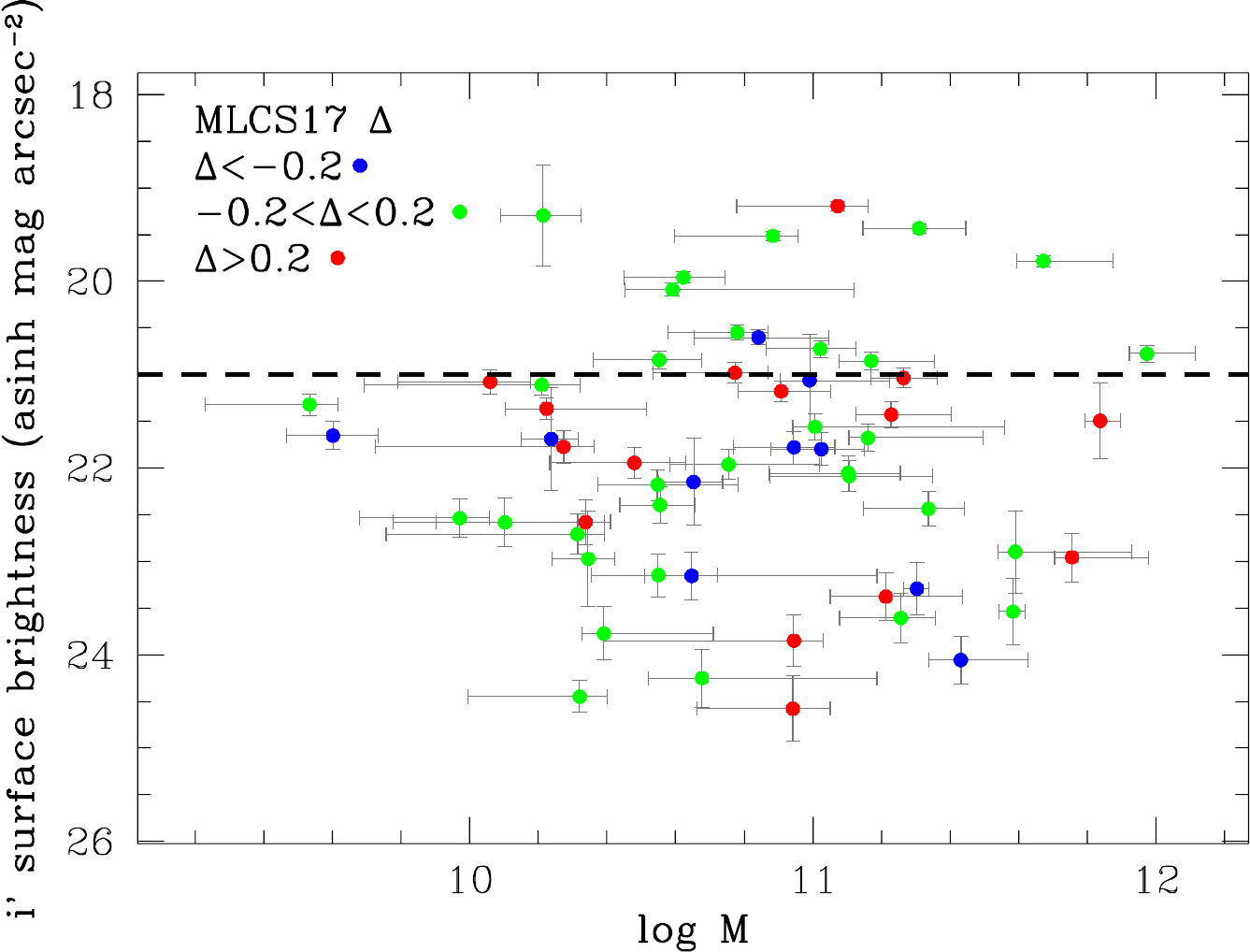}}
\hspace{0.07\textwidth}
\subfigure[]{\includegraphics[angle=0,width=3.2in]{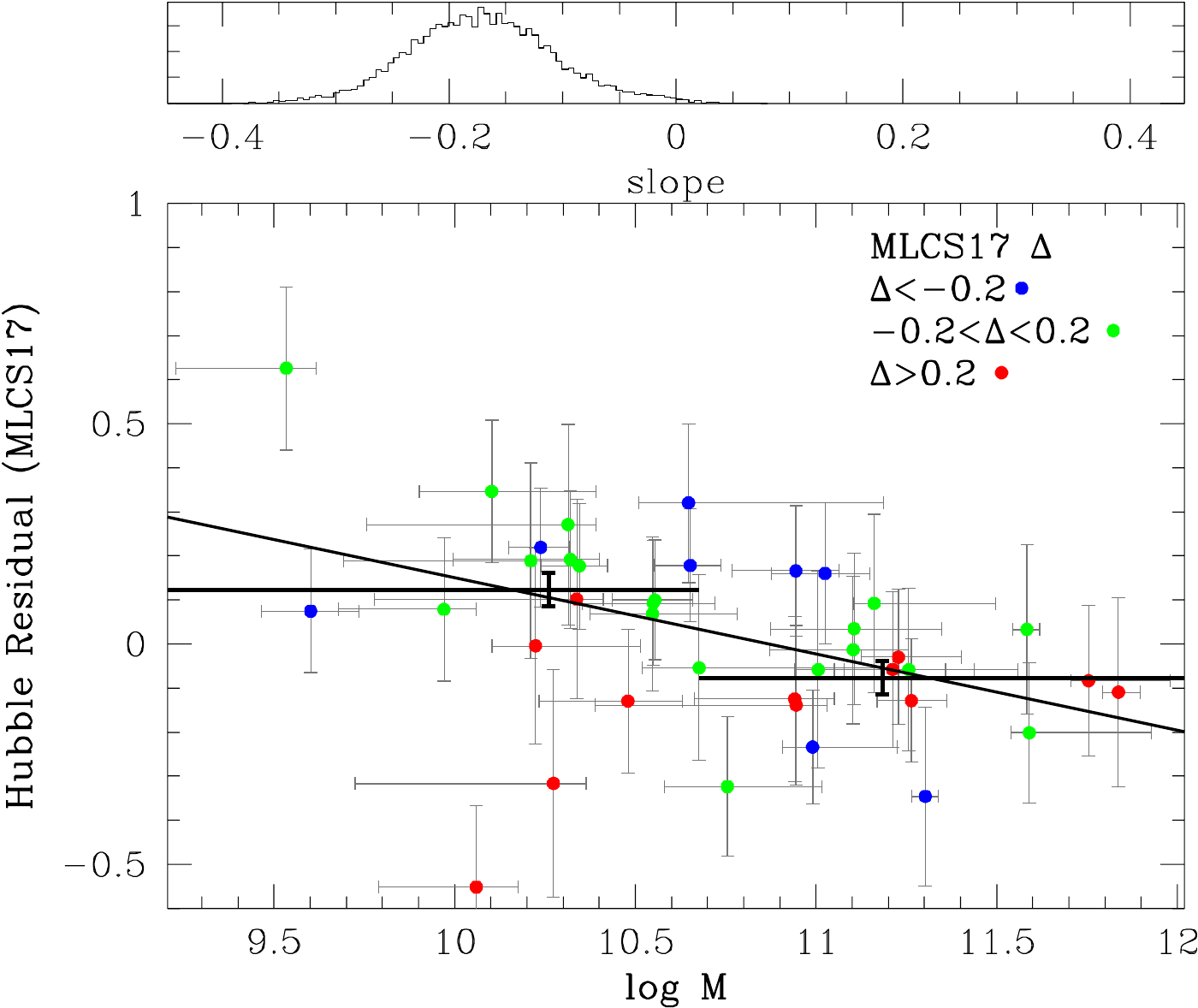}}

\caption{Hubble residual trend persists after \textit{i'}-band surface brightness cut. (a) plots the asinh magnitude \textit{i'} surface brightness in a 0.2 kpc aperture at the SN locations. We make a cut at a surface brightness of 21 asinh mag arcsecond$^{-1}$ indicated by the dashed horizontal line. The surface brightness distribution of SN locations fainter than this threshold is similar across host masses. (b) shows that the correlation between Hubble residual and host mass exists for these SN that occur in lower surface brightness locations, indicating that a Malmquist-like effect is not likely to be responsible for the observed trends in Hubble residuals with host galaxy mass. 
}
\label{fig:malmquist}
\end{figure*}

\subsection{Possible Systematic Biases}

\subsubsection{Malmquist Bias}

Photographic searches had a well-known bias against SN detection near galaxy centers where images were often saturated \citep{shaw79}. The much improved dynamic range of CCD cameras used by modern SN search programs and digital image subtraction enables detection in even high surface-brightness regions. More massive galaxies on average have regions of higher surface brightness, and we investigate whether a detection bias could explain the $\sim$0.10 mag effect associated with the galaxy mass. 
To account for a $\sim$0.10 mag trend,
SN searches would have to miss a high percentage of only slightly fainter explosions in brighter, more massive hosts.
In particular, the detection efficiency in massive hosts would have to decrease to near zero over a $\sim$10\% 
range in SN brightness. 
However, if there were such a steep drop in efficiency for brighter SN with intermediate decline rates, 
SN searches would detect few if any fast-declining SN Ia in massive galaxies because their peak intrinsic brightness can be at least 50\% fainter than the SN with intermediate decline rates detected in these galaxies. 
Because we do see SN Ia with both faster and slower decline rates in massive galaxies, it 
seems unlikely that a detection bias explains the effect with host stellar mass we observe.

To test for the possible effect of Malmquist bias, we apply a cut to remove SN Ia detected in high-surface brightness regions so that the SN in low and high mass hosts have similar distributions of galaxy surface brightness at the locations of their explosions. Figure~\ref{fig:malmquist}(a) plots the \textit{i'}-band surface brightness in a 0.2 kpc aperture at the SN Ia location. Applying a cut at a
surface brightness of 21 asinh mag arcsec$^{-1}$, shown by the dashed horizontal line, leaves surface brightness distributions that are similar for low and high mass hosts. Figure~\ref{fig:malmquist}(b) shows that a trend in Hubble residuals with host mass persists for this subset, suggesting that Malmquist bias is likely not responsible for the trend we observe. 


\subsubsection{Regional Velocity Flows}

Peculiar galaxy velocities arising from departures from the Hubble flow 
could have a systematic effect on Hubble residuals. We perform the simple test of
splitting our SN sample into lower and higher redshift groups. Both of these samples
are consistent with size and mass dependent trends.  



\subsubsection{Potential Correlations Analyzed}

The initial objective of our analysis was to constrain any correlation between the host galaxy stellar masses and Hubble residuals of our nearby SN sample to provide comparison to the \citealt{how09} SNLS study. 
We subsequently expanded our search to include potential correlations of Hubble residuals with additional host properties including galaxy color as well as the location of the SN in the host galaxy. Because galaxy properties including stellar mass, size, and color correlate with each other, the number of independent parameter searches is difficult to estimate precisely but likely is $\sim$2-3.  
The probability of finding at least one correlation with significance $p$ in an analysis of $n$ potential correlations is $p' = 1 - (1-p)^{n}$.
Accordingly, a search for correlations with 2 independent variables reduces the significance of a 2.5$\sigma$ result to $\sim$2.25$\sigma$, while a search for correlations with 3 independent variables reduces the significance of a 2.5$\sigma$ result to $\sim$2.1$\sigma$.

\section{Discussion}
\label{discuss}

Given the correlation between light curve parameters and host galaxy properties, we have been careful to investigate
whether the trends we detect with host size and mass could instead be accounted for by trends with SN Ia light curve shape or color (or dust). 
SN with a range of light curve properties are present in the high and low mass ends of the host galaxy distribution, and this allows us to disentangle potential effects of size, mass, and light curve parameters. By simultaneously
fitting for and then marginalizing over trends with light curve parameters in our fits for host-dependent trends, 
we have shown that calibration errors that correlate with light curve properties do not account for the trends we find in Hubble residuals with host galaxy size and mass.



Host galaxy metallicity, stellar age, and extinction by dust correlate with galaxy mass and may be factors responsible for 
the variation we observe in the SN Ia width-color-luminosity relation.
According to the \citealt{ti03} model, the effect of metallicity on SN Ia luminosities depends non-linearly on metallicity and would be strongest at the high metallicities of the massive galaxies in our sample.
\citealt{kas09} included the \citealt{ti03} effect in SN Ia simulations and found that increased metallicity
would alter the SN Ia width-luminosity relation so that SN Ia with intermediate to slow decline rates would 
be intrinsically brighter after light curve correction. Such a correction would be consistent with the direction of the trend we see in Hubble residuals with increasing host masses, at least for 
SN with slower decay rates. 
\citealt{kas09} did not account for the possible effects of metals on the 
dynamics of the explosion, or on the white dwarf's structure.
We note that an interesting feature of Figure~\ref{fig:size} is the apparent lack of a strong trend in the Hubble residuals with radius for SN Ia in host galaxies with sizes greater than 15 kpc. 


That the hosts of our low redshift SN Ia are substantially more massive than hosts of $0.2 < z < 0.75$ SNLS SN
helps to interpret existing studies of Hubble residuals and host galaxy properties. Figure \ref{fig:snlsvsus} shows that \citealt{how09}, analyzing SNLS SN, fit for a linear trend in Hubble residuals over a much larger range of host masses than we do and have comparatively few galaxies with masses greater than $10^{11}$ M$_{\sun}$, perhaps helping to explain the fact that they found no strong trend in Hubble residuals with host metallicity.  
\citealt{gal05} studied SN Ia in star-forming and therefore likely less massive low redshift galaxies, finding no significant trend in Hubble residuals with host metallicity. \citealt{gal08}, however, studied low redshift E/S0 galaxies, which are likely similar to the more massive hosts in our sample (four hosts overlap with our sample), and found a 98\% confidence in a trend in host metallicity.

Both \citealt{gal08} and \citealt{how09} favor trends (although the \citealt{how09} trend only has 1-$\sigma$ significance) where galaxies with higher metallicities host SN that are brighter than expected.
If the more massive galaxies in our sample have higher metallicities, the direction of the trends we find agrees with the direction of the trend in Hubble residuals with host metallicity found by \citealt{gal08}. 
The \citealt{nei09} trend in Hubble residuals with host age exists only in a subset of 
SN whose hosts have low extinction but whose light curve colors are not necessarily consistent with 
low extinction, making comparison with our trend difficult.


While we could in theory use an empirical mass-metallicity relation to test directly for the effect of metallicity in our sample following \citealt{how09} and \citealt{nei09}, it is not clear which mass-metallicity relation is appropriate. The fact that SN Ia progenitors are likely to be younger in later-type hosts may mean that their metallicities are similar to the gas-phase metallicity while SN Ia progenitors in older, early-type galaxies may track the stellar metallicity. The slope of mass-metallicity relation also becomes flatter for galaxies with approximately $10^{10.5}$ M$_{\sun}$, making mass a less sensitive proxy for metallicity.

\subsection{Strong Cosmology Implications}


Low redshift SN are essential to accurate constraints on the dark energy equation of state parameter, $w$, because they anchor the redshift-distance relation and set the expectation for the brightnesses of high redshift SN according to different cosmologies. The discovery that the universe is accelerating in its expansion originated from the observation that high redshift SN Ia were fainter than expected for a matter-only universe (\citealt{re98}; \citealt{pe99}). Low redshift SN are also used to train the MLCS2k2 light fitters and consequently influence distance measurements to high redshift SN. As Figure~\ref{fig:snlsvsus} demonstrates, only a small percentage of SN discovered by the SNLS flux-limited survey have host galaxy mass greater than $10^{10.8}$ M$_{\sun}$. However, half of the low redshift sample has hosts more massive than $10^{10.8}$ M$_{\sun}$, and this paper has shown that these SN have Hubble residuals that are on average more negative by $\sim$0.10 mag. 


To examine the sensitivity of the best-fitting value of $w$ to the host-dependent calibration error, we divide our sample into two sets according to whether the SN host has a mass greater or less than $10^{10.8}$ M$_{\sun}$. Using the wfit cosmology fitter from the Supernova Analysis package \citep{kes09b} and incorporating baryon acoustic oscillation \citep{eis05} and cosmic microwave background \citep{kom09} measurements, the best-fitting values of $w$ differ. When fit in combination with 180 higher-redshift ESSENCE, SNLS, and HigherZ SN, the 30 SN with lower mass hosts yield \mbox{\masslowcosmo} while the 30 nearby SN with higher mass hosts yield \mbox{\masshighcosmo} from MLCS17 measurements. The difference between the two estimates of $w$ is likely more significant than indicated by the overlap between their constraints, because they include 180 higher-redshift SN in common. The 30 nearby hosts with masses less than $10^{10.8}$ M$_{\sun}$ in our sample are more massive than a majority of SNLS host galaxies, so we note that we have not 
constructed a sample of SN Ia with hosts of similar masses across all redshifts.  The host-dependent shift in $w$ is greater than the statistical error on $w$ and implies that host galaxy or progenitor-dependent effects have an unappreciated and strong influence on even current SN cosmological estimates.

\section{Conclusions and Summary}

The hosts of our sample of 70 low redshift (\lowerredshift~$<z<$~\upperredshift) SN 
are substantially more massive than 
hosts of the intermediate redshift ($0.2<z<0.75$) SNLS SN.  
In our sample, 
SN Ia that occur in larger, more massive galaxies are $\sim$10\% brighter than SN Ia 
found in smaller, less massive galaxies after light curve correction 
when we calculate distance moduli using the MLCS17, MLCS31, SALT and SALT2 fitters.
When we fit simultaneously for trends with light curve parameters in addition to host galaxy properties, 
we find that these host-dependent trends
in Hubble residuals cannot be accounted for by light curve-dependent calibration error.





The variation in the SN Ia width-color-luminosity relation with host size and mass implied by these trends may reflect the correlation between galaxy mass and metallicity, increased average ages of progenitors in more massive hosts, or another environmental difference such as dust extinction.  Sensitivity of SN Ia luminosities to progenitor metallicity may become greatest at the high metallicities found in the most massive hosts in our sample \citep{ti03}.

Nearby SN Ia detected by targeted searches should be treated carefully when included in cosmology fits because their host galaxies are generally substantially
more massive than those of SN detected by flux-limited surveys at higher redshifts.  
The 30 nearby SN with host masses less than $10^{10.8}$ M$_{\sun}$ in our sample yield \mbox{\masslowcosmo} while the set of 30 nearby SN with more massive hosts yield \mbox{\masshighcosmo} with MLCS17 distances when fit in combination with 180 higher-redshift ESSENCE, SNLS, and HigherZ SN.  
By combining  information from host galaxy measurements with light curve fits, distances to SN Ia may be more accurately estimated and used to measure cosmological parameters with better control of systematics. 






\acknowledgements
 
We would like to thank especially R. Romani, M. Blanton, A. von der Linden, S. Allen, A. Mantz, C. Zhang, M. Sako, B. H{\"a}ussler, and C. Peng as well as D. Rapetti, B. Holden, D. Applegate, M. Allen, P. Behroozi, O. Ilbert, S. Arnouts, and J. Gallagher.  NSF grants AST0907903 and AST0606772 as well as the U.S. Department of Energy contract DE-AC02-70SF00515 support research on SN at Harvard University. 

Funding for the SDSS and SDSS-II has been provided by the Alfred P. Sloan Foundation, the Participating Institutions, the National Science Foundation, the U.S. Department of Energy, the National Aeronautics and Space Administration, the Japanese Monbukagakusho, the Max Planck Society, and the Higher Education Funding Council for England. 

The SDSS is managed by the Astrophysical Research Consortium for the Participating Institutions. The Participating Institutions are the American Museum of Natural History, Astrophysical Institute Potsdam, University of Basel, Cambridge University, Case Western Reserve University, University of Chicago, Drexel University, Fermilab, the Institute for Advanced Study, the Japan Participation Group, Johns Hopkins University, the Joint Institute for Nuclear Astrophysics, the Kavli Institute for Particle Astrophysics and Cosmology, the Korean Scientist Group, the Chinese Academy of Sciences (LAMOST), Los Alamos National Laboratory, the Max-Planck-Institute for Astronomy (MPIA), the Max-Planck-Institute for Astrophysics (MPA), New Mexico State University, Ohio State University, University of Pittsburgh, University of Portsmouth, Princeton University, the United States Naval Observatory, and the University of Washington.

\bibliographystyle{hapj}
\bibliography{ms}

{\it Facilities:} \facility{Sloan ()}
\begin{deluxetable*}{lcccccc}
\tablecaption{Analysis of Potential Light Curve Dependence of Hubble Residuals}
\tablecolumns{6}
\tablehead{&&\colhead{MLCS17}&\colhead{MLCS31}&\colhead{SALT}&\colhead{SALT2}&}
\startdata

&$\Delta$,$s$,$x_{1}$&\lmmseventeendelresmseventeen&\lmmthirtyonedelresmthirtyone&\lmsstrress&\lmstwostrresstwo&\\
&$A_{V}$,$c$&\lmmseventeenavresmseventeen&\lmmthirtyoneavresmthirtyone&\lmscress&\lmstwocresstwo&

\enddata
\tablecomments{Each statistic is the percentage of MCMC realizations that have slopes with opposite sign from the slope with greatest posterior probability. In parentheses is the significance of a non-zero slope followed by the number of SN included in the fit.
}
\label{tab:lightcurve}
\end{deluxetable*}

\begin{deluxetable*}{lcccccc}
\tablecaption{Magnitude Difference Between Bin Weighted Averages of Hubble Residuals}
\tablecolumns{6}
\tablehead{&&\colhead{MLCS17}&\colhead{MLCS31}&\colhead{SALT}&\colhead{SALT2}&}
\startdata

&90\% Radius&\diffnineradresmseventeen&\diffnineradresmthirtyone&\diffnineradress&\diffnineradresstwo&\\
&90\% Radius ($<$30 kpc)&\diffnineradoutlierresmseventeen&\diffnineradoutlierresmthirtyone&\diffnineradoutlierress&\diffnineradoutlierresstwo&\\
&log M ($>$9.5)&\difflogMhighresmseventeen&\difflogMhighresmthirtyone&\difflogMhighress&\difflogMhighresstwo&\\
&log M&\difflogMresmseventeen&\difflogMresmthirtyone&\difflogMress&\difflogMresstwo& \\
&log M ($>$9.5; GALEX, Neill 09)&\diffneillmasshighresmseventeen&\diffneillmasshighresmthirtyone&\diffneillmasshighress&\diffneillmasshighresstwo&\\
&log M (GALEX, Neill 09)&\diffneillmassresmseventeen&\diffneillmassresmthirtyone&\diffneillmassress&\diffneillmassresstwo&


\enddata
\tablecomments{We separate the SN into two equal-sized bins according to host property and calculate
the magnitude difference between the weighted averages of Hubble residuals of SN in each bin. 
In parentheses is the significance followed by the number of SN included in the fit.
}
\label{tab:1}
\end{deluxetable*}

\begin{deluxetable*}{lcccccc}
\tablecaption{Significance of Non-Zero Slope in Hubble Residuals}
\tablecolumns{6}
\tablehead{&&\colhead{MLCS17}&\colhead{MLCS31}&\colhead{SALT}&\colhead{SALT2}&}
\startdata

&90\% Radius&\lmnineradresmseventeen&\lmnineradresmthirtyone&\lmnineradress&\lmnineradresstwo&\\
&90\% Radius ($<$30 kpc)&\lmnineradoutlierresmseventeen&\lmnineradoutlierresmthirtyone&\lmnineradoutlierress&\lmnineradoutlierresstwo&\\
&log M ($>$9.5)&\lmlogMhighresmseventeen&\lmlogMhighresmthirtyone&\lmlogMhighress&\lmlogMhighresstwo&\\
&log M&\lmlogMresmseventeen&\lmlogMresmthirtyone&\lmlogMress&\lmlogMresstwo& \\
&log M ($>$9.5; GALEX, Neill 09)&\lmneillmasshighresmseventeen&\lmneillmasshighresmthirtyone&\lmneillmasshighress&\lmneillmasshighresstwo&\\
&log M (GALEX, Neill 09)&\lmneillmassresmseventeen&\lmneillmassresmthirtyone&\lmneillmassress&\lmneillmassresstwo&


\enddata
\tablecomments{Significance of a non-zero slope. Here we fit for the trend with host radius ($\alpha$) while holding $\beta$=$\gamma$=0.
Each statistic is the percentage of MCMC realizations that have slopes with opposite sign from the slope with greatest posterior probability.
In parentheses is the significance of a non-zero slope followed by the number of SN included in the fit.
}
\label{tab:2}
\end{deluxetable*}

\begin{deluxetable*}{lcccccc}
\tablecaption{Improvement in the Standard Deviation of Hubble Residuals after Subtracting Host-Dependent Trend}
\tablecolumns{6}
\tablehead{&&\colhead{MLCS17}&\colhead{MLCS31}&\colhead{SALT}&\colhead{SALT2}&}
\startdata

&90\% Radius&\stdbeforenineradresmseventeen~$\backslash$~\stdafternineradresmseventeen&\stdbeforenineradresmthirtyone~$\backslash$~\stdafternineradresmthirtyone&\stdbeforenineradress~$\backslash$~\stdafternineradress&\stdbeforenineradresstwo~$\backslash$~\stdafternineradresstwo&\\
&90\% Radius ($<$30 kpc)&\stdbeforenineradoutlierresmseventeen~$\backslash$~\stdafternineradoutlierresmseventeen&\stdbeforenineradoutlierresmthirtyone~$\backslash$~\stdafternineradoutlierresmthirtyone&\stdbeforenineradoutlierress~$\backslash$~\stdafternineradoutlierress&\stdbeforenineradoutlierresstwo~$\backslash$~\stdafternineradoutlierresstwo&\\
&log M&\stdbeforelogMresmseventeen~$\backslash$~\stdafterlogMresmseventeen&\stdbeforelogMresmthirtyone~$\backslash$~\stdafterlogMresmthirtyone&\stdbeforelogMress~$\backslash$~\stdafterlogMress&\stdbeforelogMresstwo~$\backslash$~\stdafterlogMresstwo& \\
&log M ($>$9.5; GALEX, Neill 09)&\stdbeforeneillmasshighresmseventeen~$\backslash$~\stdafterneillmasshighresmseventeen&\stdbeforeneillmasshighresmthirtyone~$\backslash$~\stdafterneillmasshighresmthirtyone&\stdbeforeneillmasshighress~$\backslash$~\stdafterneillmasshighress&\stdbeforeneillmasshighresstwo~$\backslash$~\stdafterneillmasshighresstwo&\\
&log M (GALEX, Neill 09)&\stdbeforeneillmassresmseventeen~$\backslash$~\stdafterneillmassresmseventeen&\stdbeforeneillmassresmthirtyone~$\backslash$~\stdafterneillmassresmthirtyone&\stdbeforeneillmassress~$\backslash$~\stdafterneillmassress&\stdbeforeneillmassresstwo~$\backslash$~\stdafterneillmassresstwo&


\enddata
\tablecomments{First value is the standard deviation in magnitudes before and the second value is the standard deviation in magnitudes after
subtracting the trend with host property, \mbox{$\alpha\times(\mbox{host property})$}, determined from fits where we hold $\beta$=$\gamma$=0. }

\label{tab:std}
\end{deluxetable*}

\begin{deluxetable*}{lcccccc}
\tablecaption{Magnitude Difference Between Bin Weighted Averages of Hubble Residuals After Removing Trends with Light Curve Parameters}
\tablecolumns{6}
\tablehead{&&\colhead{MLCS17}&\colhead{MLCS31}&\colhead{SALT}&\colhead{SALT2}&}
\startdata

&90\% Radius&\diffnineradresmseventeenmarg&\diffnineradresmthirtyonemarg&\diffnineradressmarg&\diffnineradresstwomarg&\\
&90\% Radius ($<$30 kpc)&\diffnineradoutlierresmseventeenmarg&\diffnineradoutlierresmthirtyonemarg&\diffnineradoutlierressmarg&\diffnineradoutlierresstwomarg&\\
&log M ($>$9.5)&\difflogMhighresmseventeenmarg&\difflogMhighresmthirtyonemarg&\difflogMhighressmarg&\difflogMhighresstwomarg&\\
&log M&\difflogMresmseventeenmarg&\difflogMresmthirtyonemarg&\difflogMressmarg&\difflogMresstwomarg& \\
&log M ($>$9.5; GALEX, Neill 09)&\diffneillmasshighresmseventeenmarg&\diffneillmasshighresmthirtyonemarg& \diffneillmasshighressmarg&\diffneillmasshighresstwomarg&\\
&log M (GALEX, Neill 09)&\diffneillmassresmseventeenmarg&\diffneillmassresmthirtyonemarg&\diffneillmassressmarg&\diffneillmassresstwomarg&


\enddata
\tablecomments{ Here we fit for trends with host property ($\alpha$) and light curve parameters ($\beta$,$\gamma$) simultaneously. We remove trends with light curve parameters ($\beta$,$\gamma$) from the Hubble residuals
and separate the SN into two equal-sized bins according to host property. We then calculate
the magnitude difference between the weighted averages of Hubble residuals of SN in each bin. 
In parentheses is the significance followed by the number of SN included in the fit.
}
\label{tab:3}
\end{deluxetable*}
    
\begin{deluxetable*}{lcccccc}
\tablecaption{Significance of Non-Zero Slope in Hubble Residuals After Marginalizing over Light Curve Parameters}
\tablecolumns{6}
\tablehead{&&\colhead{MLCS17}&\colhead{MLCS31}&\colhead{SALT}&\colhead{SALT2}&}
\startdata

&90\% Radius&\lmnineradresmseventeenmarg&\lmnineradresmthirtyonemarg&\lmnineradressmarg&\lmnineradresstwomarg&\\
&90\% Radius ($<$30 kpc)&\lmnineradoutlierresmseventeenmarg&\lmnineradoutlierresmthirtyonemarg&\lmnineradoutlierressmarg&\lmnineradoutlierresstwomarg&\\
&log M ($>$9.5)&\lmlogMhighresmseventeenmarg&\lmlogMhighresmthirtyonemarg&\lmlogMhighressmarg&\lmlogMhighresstwomarg&\\
&log M&\lmlogMresmseventeenmarg&\lmlogMresmthirtyonemarg&\lmlogMressmarg&\lmlogMresstwomarg& \\
&log M ($>$9.5; GALEX, Neill 09)&\lmneillmasshighresmseventeenmarg&\lmneillmasshighresmthirtyonemarg&\lmneillmasshighressmarg&\lmneillmasshighresstwomarg&\\
&log M (GALEX, Neill 09)&\lmneillmassresmseventeenmarg&\lmneillmassresmthirtyonemarg&\lmneillmassressmarg&\lmneillmassresstwomarg&

\enddata
\tablecomments{Significance of a non-zero slope. Here we have fit for trends with host property ($\alpha$) and light curve parameters ($\beta$,$\gamma$) simultaneously. Each statistic is the percentage of MCMC realizations that have slopes with opposite sign from the slope with greatest posterior probability. 
In parentheses is the significance of a non-zero slope followed by the number of SN included in the fit.
}
\label{tab:4}
\end{deluxetable*}

\begin{deluxetable*}{lcccccc}
\tablecaption{Improvement in the Standard Deviation of Hubble Residuals after Subtracting Host-Dependent Trend from Simultaneous Fit}
\tablecolumns{6}
\tablehead{&&\colhead{MLCS17}&\colhead{MLCS31}&\colhead{SALT}&\colhead{SALT2}&}
\startdata

&90\% Radius&\stdbeforenineradresmseventeenmarg~$\backslash$~\stdafternineradresmseventeenmarg&\stdbeforenineradresmthirtyonemarg~$\backslash$~\stdafternineradresmthirtyonemarg&\stdbeforenineradressmarg~$\backslash$~\stdafternineradressmarg&\stdbeforenineradresstwomarg~$\backslash$~\stdafternineradresstwomarg&\\
&90\% Radius ($<$30 kpc)&\stdbeforenineradoutlierresmseventeenmarg~$\backslash$~\stdafternineradoutlierresmseventeenmarg&\stdbeforenineradoutlierresmthirtyonemarg~$\backslash$~\stdafternineradoutlierresmthirtyonemarg&\stdbeforenineradoutlierressmarg~$\backslash$~\stdafternineradoutlierressmarg&\stdbeforenineradoutlierresstwomarg~$\backslash$~\stdafternineradoutlierresstwomarg&\\
&log M&\stdbeforelogMresmseventeenmarg~$\backslash$~\stdafterlogMresmseventeenmarg&\stdbeforelogMresmthirtyonemarg~$\backslash$~\stdafterlogMresmthirtyonemarg&\stdbeforelogMressmarg~$\backslash$~\stdafterlogMressmarg&\stdbeforelogMresstwomarg~$\backslash$~\stdafterlogMresstwomarg& \\
&log M ($>$9.5; GALEX, Neill 09)&\stdbeforeneillmasshighresmseventeenmarg~$\backslash$~\stdafterneillmasshighresmseventeenmarg&\stdbeforeneillmasshighresmthirtyonemarg~$\backslash$~\stdafterneillmasshighresmthirtyonemarg&\stdbeforeneillmasshighressmarg~$\backslash$~\stdafterneillmasshighressmarg&\stdbeforeneillmasshighresstwomarg~$\backslash$~\stdafterneillmasshighresstwomarg&\\
&log M (GALEX, Neill 09)&\stdbeforeneillmassresmseventeenmarg~$\backslash$~\stdafterneillmassresmseventeenmarg&\stdbeforeneillmassresmthirtyonemarg~$\backslash$~\stdafterneillmassresmthirtyonemarg&\stdbeforeneillmassressmarg~$\backslash$~\stdafterneillmassressmarg&\stdbeforeneillmassresstwomarg~$\backslash$~\stdafterneillmassresstwomarg&


\enddata

\tablecomments{First value is the standard deviation in magnitudes before and the second value is the standard deviation in magnitudes after
subtracting the trend with host property, \mbox{$\alpha\times(\mbox{host property})$}, but not the trends with light curve properties ($\beta$,$\gamma$). Here the trend with host property ($\alpha$) is determined in fits where we simultaneously fit for trends with light curve parameters ($\beta$,$\gamma$). }

\label{tab:stdmarg}
\end{deluxetable*}

\begin{deluxetable*}{lcccccccc}
\tablecaption{Bayesian Information Criteria}
\tablecolumns{7}
\tablehead{&Parameters&Host Property&Light Curve&\colhead{MLCS17}&\colhead{MLCS31}&\colhead{SALT}&\colhead{SALT2}&}
\startdata

&$\delta$&...&$...$&\flatresmseventeenmargbic&\flatresmthirtyonemargbic&\flatressmargbic&\flatresstwomargbic&\\
&$\beta,\delta$&...&$\Delta/s/x_{1}$&\mseventeendelakresmseventeenbic&\mthirtyonedelakresmthirtyonebic&\sstrakressbic&\stwostrakresstwobic&\\
&$\alpha,\beta,\delta$&90\% Radius&$\Delta/s/x_{1}$&\nineradakresmseventeenmargpartbic&\nineradakresmthirtyonemargpartbic&\nineradakressmargpartbic&\nineradakresstwomargpartbic&\\
&$\alpha,\beta,\delta$&log M ($>$9.5)&$\Delta/s/x_{1}$&\logmhighakresmseventeenmargpartbic&\logmhighakresmthirtyonemargpartbic&\logmhighakressmargpartbic&\logmhighakresstwomargpartbic&\\
&$\alpha,\beta,\delta$&log M ($>$9.5; GALEX, Neill 09)&$\Delta/s/x_{1}$&\neillmasshighakresmseventeenmargpartbic&\neillmasshighakresmthirtyonemargpartbic&\neillmasshighakressmargpartbic&\neillmasshighakresstwomargpartbic&

\enddata
\tablecomments{BIC increments after adding linear terms ($\beta$,$\alpha$) to a constant model ($\delta$) of Hubble residuals. Only differences between BIC values are meaningful, and models with lower values are favored. A difference between 0 and 2 is weak evidence, 2 and 6 evidence, and greater than 6 strong evidence against the model with higher BIC. 
The improvements in $\chi^{2}$ provide evidence for the addition of trends with host size for all light curve fitter
and with host mass for MLCS17, MLCS31, and SALT2. 
An $A_{V}/c$ trend ($\gamma$) is not included in this analysis because it has a slope consistent with zero and produces no significant reduction in the fit $\chi^{2}$.  }
\label{tab:bic}
\end{deluxetable*}


\begin{deluxetable*}{lcccccccc}
\tablecaption{Corrected Akaike Information Criteria}
\tablecolumns{7}
\tablehead{&Parameters&Host Property&Light Curve&\colhead{MLCS17}&\colhead{MLCS31}&\colhead{SALT}&\colhead{SALT2}&}
\startdata

&$\delta$&...&$...$&\flatresmseventeenmargakaike&\flatresmthirtyonemargakaike&\flatressmargakaike&\flatresstwomargakaike&\\
&$\beta,\delta$&...&$\Delta/s/x_{1}$&\mseventeendelakresmseventeenakaike&\mthirtyonedelakresmthirtyoneakaike&\sstrakressakaike&\stwostrakresstwoakaike&\\
&$\alpha,\beta,\delta$&90\% Radius&$\Delta/s/x_{1}$&\nineradakresmseventeenmargpartakaike&\nineradakresmthirtyonemargpartakaike&\nineradakressmargpartakaike&\nineradakresstwomargpartakaike&\\
&$\alpha,\beta,\delta$&log M ($>$9.5)&$\Delta/s/x_{1}$&\logmhighakresmseventeenmargpartakaike&\logmhighakresmthirtyonemargpartakaike&\logmhighakressmargpartakaike&\logmhighakresstwomargpartakaike&\\
&$\alpha,\beta,\delta$&log M ($>$9.5; GALEX, Neill 09)&$\Delta/s/x_{1}$&\neillmasshighakresmseventeenmargpartakaike&\neillmasshighakresmthirtyonemargpartakaike&\neillmasshighakressmargpartakaike&\neillmasshighakresstwomargpartakaike&

\enddata
\tablecomments{AIC$_{c}$ increments after adding linear terms ($\beta$,$\alpha$) to a constant model ($\delta$) of Hubble residuals. 
The improvements in $\chi^{2}$ provide evidence for the addition of trends with host properties ($\alpha$). $\Delta$AIC$_{c}$ values are generally more negative than $\Delta$BIC values because BIC includes a stronger penalty for additional fit parameters.
An $A_{V}/c$ trend ($\gamma$) is not included in this analysis because it has a slope consistent with zero and produces no significant reduction in the fit $\chi^{2}$.  }
\label{tab:aic}
\end{deluxetable*}

\clearpage

\LongTables

\begin{deluxetable}{lcccccc}
\tablecaption{Nearby Host Measurements}
\tablewidth{0pt}
\tablecolumns{6}
\tablehead{\colhead{SN}&\colhead{z$_{CMB}$}&\colhead{90\% \textit{i}-band radius}&\colhead{M$^{*}-$}&\colhead{M$^{*}$}&\colhead{M$^{*}+$}\\
\colhead{ }&\colhead{ }&\colhead{(kpc) }&\colhead{ }&\colhead{(log M$_{\sun}$) }&\colhead{ }}
\startdata
1992P & 0.0265 & 9.63 & 10.31 & 10.55 & 10.72 & \\
1993ac & 0.0489 & 22.32 & 11.55 & 11.58 & 11.62 & \\
1994M & 0.0243 & 27.10 & 10.99 & 11.21 & 11.37 & \\
1994S & 0.0152 & 10.51 & 10.17 & 10.68 & 10.83 & \\
1994T & 0.0357 & 8.15 & 10.11 & 9.74 & 10.74 & \\
1995ac & 0.0488 & 17.43 & 10.76 & 10.99 & 11.08 & \\
1996C & 0.0275 & 10.02 & 9.81 & 10.10 & 10.30 & \\
1997Y & 0.0166 & 8.74 & 10.06 & 10.59 & 10.73 & \\
1997cn & 0.0170 & 16.73 & 11.52 & 11.56 & 11.60 & \\
1998ab & 0.0279 & 14.58 & 10.49 & 10.75 & 10.93 & \\
1999cc & 0.0315 & 21.47 & 11.05 & 11.23 & 11.33 & \\
1999gd & 0.0193 & 13.60 & 9.89 & 10.02 & 10.54 & \\
2000ce & 0.0165 & 11.79 & 10.19 & 10.72 & 10.85 & \\
2001N & 0.0221 & 15.01 & 10.81 & 10.88 & 11.17 & \\
2001V & 0.0162 & 16.65 & 11.27 & 11.30 & 11.34 & \\
2001ah & 0.0583 & 29.36 & 10.86 & 11.01 & 11.08 & \\
2001da & 0.0160 & 8.48 & 10.10 & 10.21 & 10.73 & \\
2001en & 0.0155 & 17.27 & 10.45 & 11.01 & 11.07 & \\
2001gb & 0.0268 & 12.61 & 10.42 & 10.53 & 10.81 & \\
2001ic & 0.0429 & 29.91 & 11.78 & 11.84 & 11.88 & \\
2001ie & 0.0312 & 15.00 & 11.17 & 11.26 & 11.44 & \\
2002G & 0.0346 & 9.76 & 10.76 & 10.91 & 11.03 & \\
2002bf & 0.0249 & 10.51 & 10.64 & 10.84 & 11.03 & \\
2002bz & 0.0377 & 25.09 & 10.98 & 11.17 & 11.26 & \\
2002ck & 0.0302 & 22.08 & 10.94 & 11.10 & 11.34 & \\
2002hw & 0.0163 & 12.91 & 10.01 & 10.55 & 10.67 & \\
2002jy & 0.0201 & 19.50 & 10.11 & 10.65 & 10.78 & \\
2003U & 0.0279 & 17.93 & 10.33 & 10.48 & 10.73 & \\
2003cq & 0.0337 & 25.75 & 11.23 & 11.34 & 11.53 & \\
2003ic & 0.0542 & 54.79 & 11.83 & 11.97 & 12.03 & \\
2004L & 0.0334 & 13.39 & 10.50 & 10.62 & 10.80 & \\
2005eq & 0.0284 & 17.15 & 10.82 & 10.94 & 11.12 & \\
2005hc & 0.0450 & 12.42 & 10.45 & 10.55 & 10.67 & \\
2005hj & 0.0570 & 4.17 & 9.50 & 9.63 & 9.77 & \\
2005ir & 0.0753 & 8.38 & 10.16 & 10.24 & 10.33 & \\
2005ki & 0.0208 & 26.98 & 10.86 & 10.95 & 11.50 & \\
2005mc & 0.0261 & 18.58 & 10.87 & 10.97 & 11.26 & \\
2005ms & 0.0259 & 11.37 & 10.24 & 10.32 & 10.65 & \\
2006S & 0.0329 & 15.05 & 10.45 & 10.60 & 10.76 & \\
2006ac & 0.0236 & 16.43 & 10.86 & 11.11 & 11.34 & \\
2006ak & 0.0391 & 10.34 & 10.69 & 10.78 & 10.98 & \\
2006al & 0.0691 & 5.63 & 10.27 & 10.35 & 10.45 & \\
2006an & 0.0651 & 2.62 & 7.39 & 7.68 & 7.94 & \\
2006ar & 0.0229 & 6.73 & 9.93 & 10.22 & 10.35 & \\
2006az & 0.0316 & 18.46 & 11.48 & 11.67 & 11.75 & \\
2006bd & 0.0265 & 21.49 & 10.95 & 11.23 & 11.32 & \\
2006bq & 0.0215 & 12.77 & 11.00 & 11.10 & 11.38 & \\
2006bt & 0.0325 & 16.11 & 11.23 & 11.43 & 11.52 & \\
2006bz & 0.0277 & 7.24 & 10.27 & 10.56 & 10.65 & \\
2006cc & 0.0328 & 12.09 & 10.83 & 11.04 & 11.12 & \\
2006cf & 0.0423 & 17.97 & 10.92 & 11.02 & 11.18 & \\
2006cg & 0.0288 & 5.43 & 9.94 & 10.06 & 10.33 & \\
2006cj & 0.0684 & 9.64 & 10.57 & 10.65 & 10.75 & \\
2006cp & 0.0233 & 15.15 & 10.07 & 10.39 & 10.45 & \\
2006cq & 0.0491 & 20.87 & 10.90 & 11.03 & 11.17 & \\
2006cs & 0.0241 & 10.17 & 10.89 & 11.23 & 11.28 & \\
2006ef & 0.0170 & 8.95 & 10.24 & 10.31 & 10.87 & \\
2006ej & 0.0192 & 6.25 & 10.83 & 11.16 & 11.22 & \\
2006nz & 0.0372 & 2.90 & 10.03 & 10.12 & 10.31 & \\
2006oa & 0.0589 & 8.89 & 8.80 & 8.96 & 9.09 & \\
2006ob & 0.0582 & 28.33 & 11.14 & 11.24 & 11.35 & \\
2006on & 0.0688 & 4.95 & 10.14 & 10.23 & 10.35 & \\
2006or & 0.0217 & 17.90 & 11.01 & 11.21 & 11.43 & \\
2006te & 0.0321 & 12.48 & 10.43 & 10.55 & 10.74 & \\
2007F & 0.0242 & 12.22 & 9.88 & 9.97 & 10.26 & \\
2007R & 0.0296 & 16.06 & 11.17 & 11.31 & 11.47 & \\
2007ap & 0.0162 & 7.55 & 10.24 & 10.32 & 10.87 & \\
2007ar & 0.0534 & 25.75 & 11.61 & 11.65 & 11.69 & \\
2007ba & 0.0391 & 18.99 & 11.09 & 11.19 & 11.37 & \\
2007bc & 0.0219 & 14.86 & 10.83 & 10.94 & 11.24 & \\
2007bd & 0.0320 & 14.01 & 10.68 & 10.77 & 11.01 & \\
2007bz & 0.0227 & 5.66 & 9.45 & 9.53 & 9.84 & \\
2007ci & 0.0191 & 15.42 & 10.63 & 10.72 & 11.28 & \\
2008L & 0.0187 & 7.32 & 10.18 & 10.27 & 10.82 & \\
2008af & 0.0346 & 29.23 & 11.53 & 11.76 & 11.81 & \\
2008bf & 0.0253 & 20.39 & 11.25 & 11.59 & 11.64 & 
\enddata
\label{tab:all}
\end{deluxetable}

\end{document}